\documentclass[aps,prc,reprint,superscriptaddress,showpacs,amsmath,amssymb,longbibliography,lengthcheck]{revtex4-1}
\usepackage{graphicx}
\usepackage{mathrsfs}
\usepackage{txfonts}
\usepackage{hyperref}

\begin{document}

\title{Systematic investigations of deep sub-barrier fusion reactions
using an adiabatic approach}

\author{Takatoshi Ichikawa}%
\affiliation{Yukawa Institute for Theoretical Physics, Kyoto University,
Kyoto 606-8502, Japan}
\date{\today}

\begin{abstract}
\begin{description}
 \item[Background] At extremely low incident energies, unexpected
	    decreases in fusion cross sections, compared to the standard
	    coupled-channels (CC) calculations, have been observed in a
	    wide range of fusion reactions. These significant reductions
	    of the fusion cross sections are often referred to as the
	    fusion hindrance. However, the physical origin of the fusion
	    hindrance is still unclear.
 \item[Purpose] To describe the fusion hindrance based on an adiabatic
	    approach, I propose a novel extension of the standard CC
	    model by introducing a damping factor that describes a
	    smooth transition from sudden to adiabatic processes.  I
	    demonstrate the performance of this model by systematically
	    investigating various deep sub-barrier fusion reactions.
 \item[Method] I extend the standard CC model by introducing a damping
	    factor into the coupling matrix elements in the standard CC
	    model. This avoids double counting of the CC effects, when
	    two colliding nuclei overlap one another. I adopt the
	    Yukawa-plus-exponential (YPE) model as a basic heavy ion-ion
	    potential, which is advantageous for a unified description
	    of the one- and two-body potentials. For the purpose of
	    these systematic investigations, I approximate the one-body
	    potential with a third-order polynomial function based on
	    the YPE model.
 \item[Results] Calculated fusion cross sections for the medium-heavy
	    mass systems of $^{64}$Ni + $^{64}$Ni, $^{58}$Ni +
	    $^{58}$Ni, and $^{58}$Ni + $^{54}$Fe, the medium-light mass
	    systems of $^{40}$Ca + $^{40}$Ca, $^{48}$Ca + $^{48}$Ca, and
	    $^{24}$Mg + $^{30}$Si, and the mass-asymmetric systems of
	    $^{48}$Ca + $^{96}$Zr and $^{16}$O + $^{208}$Pb are
	    consistent with the experimental data. The astrophysical S
	    factor and logarithmic derivative representations of these
	    are also in good agreement with the experimental data. The
	    values obtained for the individual radius and diffuseness
	    parameters in the damping factor, which reproduce the fusion
	    cross sections well, are nearly equal to the average value
	    for all the systems.
 \item[Conclusions] Since the results calculated with the damping factor
	    are in excellent agreement with the experimental data in all
	    systems, I conclude that the smooth transition from the
	    sudden to adiabatic processes occurs and that a
	    coordinate-dependent coupling strength is responsible for
	    the fusion hindrance. In all systems, the potential energies
	    at the touching point $V_{\rm Touch}$ strongly correlate
	    with the incident threshold energies for which the fusion
	    hindrance starts to emerge, except for the medium-light mass
	    systems.
 \end{description}
\end{abstract}

\pacs{24.10.Eq, 25.70.Jj}
\keywords{}

\maketitle
\section{Introduction}

Heavy-ion fusion reactions are an important probe to investigate the
fundamental features of the macroscopic tunneling for many-body quantum
systems.  The Coulomb barrier is formed when a projectile approaches a
target, because of the strong cancellation between the Coulomb repulsion
and nuclear attractive forces.  Fusion occurs when the projectile
penetrates through this Coulomb barrier.  The fusion reaction at
incident energies around the Coulomb barrier is called the sub-barrier
fusion reaction.  An important observation of the sub-barrier fusion
reactions is that the measured fusion cross sections exhibit strong
enhancements compared to estimations using a simple one-dimensional
model \cite{DHRS98}. These enhancements have been accounted for in terms
of strong couplings between the relative motion of colliding nuclei and
the intrinsic degrees of freedom, such as the collective vibrations of
the target and/or projectile.  In this context, the coupled-channels
(CC) model based on this picture has been successful in describing these
enhancements \cite{Bal98,Hagino12}.

Because of recent progress in experimental techniques, it is possible to
precisely measure the fusion cross sections down to extremely low
incident energies, the so-called ``deep sub-barrier energies''.  The
experimental data revealed that significant decreases in the fusion
cross sections at deep sub-barrier energies, compared to the standard CC
calculations, emerge in a wide range of reaction systems
\cite{Jiang02,Jiang04,Jiang06} (see Ref.~\cite{Back14} for details).
These significant decreases in the fusion cross section are often
referred to as fusion hindrance.  Below, the incident energy for which
the fusion hindrance starts to emerge is referred to as ``the incident
threshold energy for fusion hindrance''.

A key quantity for understanding the fusion hindrance is the potential
energy at the touching point $V_{\rm Touch}$, which strongly correlates
with the incident threshold energy for fusion hindrance \cite{ich07-2}.
When the incident energy is lower than $V_{\rm Touch}$, the inner
turning point of the potential energy becomes inside the touching point
(see Fig.~\ref{spic}). Namely, the projectile is still in a classically
forbidden region when the two colliding nuclei touch with each other.
As a result, the colliding nuclei must penetrate through a residual
barrier with an overlapping configuration before fusion occurs.  Thus,
the fusion hindrance would be associated with the dynamics in the
overlapping region of the two colliding nuclei.

To describe the fusion hindrance associated with the dynamics in the
overlapping region, two approaches with different assumptions that
contradict one another have been proposed \cite{Back14}. One is the
sudden approach proposed by Mi\c{s}icu and Esbensen \cite{mis06}, which
assumes that fusion occurs rapidly. They considered the Pauli principle
effect that acts when two colliding nuclei overlap one another. Thus,
they constructed a heavy ion-ion potential with a shallow potential
pocket based on the frozen-density approximation. They systematically
investigated various reaction systems to test the performance of the
sudden model \cite{Esb07,Esb10,mis11,Monta12,Jiang14}.

The other is the adiabatic approach proposed by Ichikawa {\it et
al.}~\cite{ich07-1,ich09}, which assumes that fusion slowly occurs and
that neck formation between two colliding nuclei occurs in the
overlapping region.  In these considerations, the sudden and adiabatic
processes are smoothly jointed by phenomenologically introducing a
damping factor in the coupling form factor to avoid double counting of
the CC effects.  Later, it was shown that the physical origin of this
damping factor is the damping of quantum vibrations near the touching
point of two colliding nuclei using the random-phase approximation (RPA)
method for the $^{16}$O + $^{16}$O, $^{40}$Ca + $^{40}$Ca, and $^{16}$O
+$^{208}$Pb systems \cite{ich13,ich15}.

Another approach, different from the CC model, has recently been
developed to describe heavy-ion fusion reactions based on
self-consistent mean-field theory \cite{Back14}. In this approach, the
time-dependent Hatree-Fock (TDHF) method is often used to extract an
adiabatic heavy ion-ion potential. Umar and Oberacker proposed the
density-constrained TDHF method and demonstrated the energy-dependent
heavy ion-ion potential with inertia mass relative to the center-of-mass
distance between two colliding nuclei
\cite{Umar06,Umar09,Umar14,Keser12}. Although in this approach, any
input parameters are not required to calculate fusion cross sections
once one determines an energy-density functional, its mechanism for the
fusion hindrance is still unclear.

In this paper, I systematically investigate the fusion hindrance using
the adiabatic approach to test the model's performance in various
reaction systems.  Later, I show that the adiabatic approach works very
well in many systems, strongly indicating that indeed the smooth
transition from sudden to adiabatic processes occurs in deep sub-barrier
fusion reactions, i.e.~this coordinate-dependent coupling strength is
responsible for the fusion hindrance.  I also show that a difference
between the adiabatic and sudden models appears in the average angular
momentum of compound nuclei. I also discuss the strong correlation
between $V_{\rm Touch}$ and the incident threshold energy for 
fusion hindrance.

The rest of the paper is organized as follows. In Section II, I describe
the theoretical framework and how to construct a heavy ion-ion
potential.  In Section III, I present the results of the systematic
calculations by applying the adiabatic approach for the medium-heavy,
medium-light, and mass-asymmetric systems.  In Section IV, I discuss the
average angular momentum of compound nuclei calculated with the
adiabatic and sudden models, and the correlation between $V_{\rm Touch}$
and the threshold incident energy for activating the fusion hindrance.
I summarize my studies in Section V.

\section{THEORETICAL FRAMEWORK}
\subsection{Concept of a smooth transition from sudden to   
  adiabatic processes}

\begin{figure}[htbp]
 \includegraphics[keepaspectratio,width=7cm]{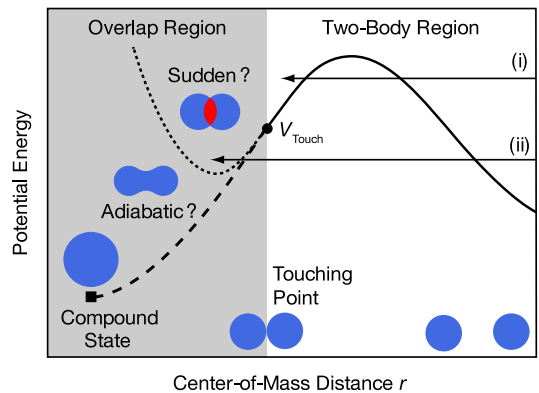} \caption{(Color
 online) Schematic picture of a heavy ion-ion potential versus the
 center-of-mass distance $r$ between colliding nuclei. The solid
 circle and square indicate the touching point of the colliding nuclei
 and its compound state, respectively. The gray area represents the
 overlapping region of the colliding nuclei. The dashed and dotted lines
 indicate the potential energy curves for the adiabatic and sudden
 processes, respectively. } \label{spic}
\end{figure}

Here, I discuss how to describe the fusion hindrance based on the
adiabatic approach and show a concept of smooth transition from 
sudden to adiabatic processes.

An important aspect of fusion reactions at deep sub-barrier incident
energies is that the inner turning point of a heavy ion-ion potential
would be located deep within the touching point of the colliding nuclei.
Figure \ref{spic} shows a schematic picture of a heavy ion-ion potential
versus incident energies in a fusion reaction.  The solid line indicates
a potential energy curve in the two-body region.  The solid circle
indicates the potential energy at the touching point $V_{\rm Touch}$ of
the colliding nuclei.  The gray area represents the overlapping region
of the colliding nuclei.  At incident energies around the Coulomb
barrier, the inner turning point is still far outside of the touching
point [line (i) in Fig.~\ref{spic}].  At these energies, one usually
assumes that a compound nucleus is automatically formed once the
projectile penetrates the Coulomb barrier because of strong nuclear
attractive forces in the classically allowed region.  On the other hand,
at incident energies below $V_{\rm Touch}$, the inner turning point
appears more deeply within the touching point [line (ii) in
Fig.~\ref{spic}]. Namely, the projectile is still in the classically
forbidden region when the colliding nuclei touch one another.

After touching, a composite system is formed, which evolves in the
classically forbidden region toward its compound state by overlapping
between projectile- and target-like nuclei.  Since this involves the
penetration of a residual Coulomb barrier, fusion cross sections are
naturally hindered by the tunneling factor.  In Ref.~\cite{ich07-2}, a
strong correlation between $V_{\rm Touch}$ and the incident threshold
energy for fusion hindrance is found by systematically
investigating various experimental data.  Thus, the dynamics after the
nuclei collide plays an essential role in significantly decreasing the
fusion cross section at deep sub-barrier incident energies.

A description for the evolution toward the compound state strongly
depends on which model is employed in the overlapping region.  There are
mainly two assumptions that contradict one another.  One is the sudden
approach, where fusion rapidly occurs. Here, the potential energy curve
would have a shallow potential pocket due to the strong overlapping of
the two colliding nuclei [dotted line in Fig.~\ref{spic}]. The other is
the adiabatic approach, where the density distribution of the composite
system evolves with the lowest energy configuration [dashed line in
Fig.~\ref{spic}]. In this paper, I focus on applying the adiabatic
process to the standard CC model framework.

However, one cannot directly apply the adiabatic potential calculated
with the lowest energy configuration to the standard CC model, because
its direct application leads to double counting of the CC
effects. Channel coupling already includes many effects of the adiabatic
process, including neck formation between the colliding nuclei.  To
avoid such double counting, I developed a full quantum mechanical model,
whereby the CC effect in the two-body system is smoothly jointed to the
adiabatic potential tunneling for the one-body system.

In the CC calculations, one often employs the incoming wave boundary
condition in order to simulate a compound nucleus formation.  To
construct an adiabatic potential model with it, I assume the following
conditions: (1) before the target and projectile touch one another, the
standard CC model in the two-body system works well; (2) after the
target and projectile appreciably overlap one another, the fusion
process is governed by a single adiabatic one-body potential, whereby
the excitation on the adiabatic base is neglected; and (3) the
transition from the two- to one-body treatments occurs near the touching
point, where all physical quantities are smoothly joined.  These are the
important conditions, which should be taken into account in the
adiabatic approach under the framework of the CC model.

  \subsection{An extension of the standard coupled-channel model}

Before describing an extension of the standard CC model, taking into
account the adiabatic process, I first briefly describe the standard CC
model (for details see Refs.~\cite{Esb87,ccfull,Hagino12}).

For heavy-ion fusion reactions, the no-Coriolis approximation is often
used~\cite{ccfull,Hagino12}. Here, one can replace the angular momentum
of the relative motion of colliding nuclei in each channel by the total
angular momentum, $J$.  For simplicity, the index $J$ is suppressed and
a simplified notation $n =\{\alpha,\ell ,I \}$ is used in the following,
where $\alpha$ denotes any quantum numbers separate from the angular
momenta, and $\ell$ and $I$ denote the orbital and intrinsic angular
momenta, respectively. The CC equations \cite{ccfull,Hagino12} are then
given by
\begin{multline}
\Biggl[-\frac{\hbar^2}{2\mu}\frac{d^2}{dr^2}
 +\frac{J(J+1)\hbar^2}{2\mu r^2}+V^{(0)}(r)
 +\epsilon_n-E\Biggr]u_n(r) \\ +\sum_m V_{nm}(r)u_m(r) = 0, \label{cceq}
\end{multline}
where $r$ is the radial component of the relative motion coordinate,
$\mu$ is the reduced mass, $E$ is the incident energy in the
center-of-mass frame, and $\epsilon_n$ is the excitation energy of the
$n$-th channel. A bare nuclear potential $V^{(0)}$ consisting of the
Coulomb and nuclear interactions is given by $V^{(0)}(r)=Z_TZ_Pe^2/r +
V_N^{(0)}(r)$. The matrix elements of the coupling Hamiltonian $V_{nm}$
are calculated with the collective model including the Coulomb and
nuclear components.

The CC equations are solved by imposing the incoming wave boundary
condition \cite{ccfull,Hagino12} at the minimum of the potential
pocket $r_{\rm min}$ inside the Coulomb barrier. This condition is
expressed as
\begin{alignat}{2}
 u_n(r) &\sim T_n\exp\left(-i\int_{r_{\rm min}}^rk_n(r')dr'\right)&& (r \le r_{\rm min}) \\
 &=H_{J}^{(-)}(k_nr)\delta_{n,0} + S_nH_{J}^{(+)}(k_nr) &\quad&(r \rightarrow \infty),
 \label{cwf}
\end{alignat}
where $S_n$ are the $S$ matrix, $T_n$ are the transmission coefficients,
and $H^{(+)}$ and $H^{(-)}$ are the outgoing and incoming Coulomb wave
functions, respectively.
The local wave number for the $n$-th channel $k_n$(r) is given by 
\begin{equation}
 k_n(r)=\sqrt{\frac{2\mu}{\hbar^2}\left(E-\epsilon_n-\frac{J(J+1)\hbar^2}{2\mu r^2}-V^{(0)}(r)-V_{nm}(r)\right)}.
\end{equation}
By taking a summation over all possible intrinsic states,
the inclusive penetrability is given by
\begin{equation}
 P_J(E)=\sum_n\frac{k_n(r_{\rm min})}{k_0}|T_n|^2,
\end{equation}
where $k_n=k_n(r=\infty)$ and the ground state of the target nucleus is
denoted by $n=0$.  The fusion cross section $\sigma_{\rm fus}$ is thus
obtained as
\begin{equation}
 \sigma_{\rm fus}(E)=\frac{\pi}{k^2}\sum_J(2J + 1)P_J(E).
\end{equation}

In coupling matrix elements, I consider only vibrational couplings in
this paper.  The nuclear coupling Hamiltonian can be generated by
changing the target radius in the nuclear potential of $V_N^{(0)}$ to a
dynamical operator $R_0\rightarrow R_0 + \hat{O}_\lambda$. Therefore,
the nuclear coupling Hamiltonian is given by
$V_N(r,\hat{O}_\lambda)=V_N^{(0)}(r-\hat{O}_\lambda)$
\cite{ccfull,Hagino12}.  For the vibrational coupling, the operator
$\hat{O}_\lambda$ is given by $\hat{O}_\lambda=\beta_\lambda /
\sqrt{4\pi}\cdot R_T(\alpha_{\lambda 0}^\dagger + \alpha_{\lambda 0})$,
where $\alpha_{\lambda 0}^\dagger$ and $\alpha_{\lambda 0}$ are the
creation and annihilation operators of the phonons, respectively, and
$R_T$ is the radius of the target nucleus.  Here, the eigenvalues
$\lambda_\alpha$ and the eigenvectors $|\alpha\rangle$ of
$\hat{O}_\lambda$ are given by
$\hat{O_\lambda}|\alpha\rangle=\lambda_\alpha|\alpha\rangle$.  The
deformation parameter $\beta_\lambda$ is an input parameter and can be
estimated from an experimental transition probability $B(E\lambda)$
which is given by
\begin{equation}
\beta_\lambda =\frac{4\pi}{3Z_TR_T^\lambda}\sqrt{\frac{B(E\lambda)\uparrow}{e^2}},
 \label{beta}
\end{equation}
where $Z_T$ is the proton number and $e$ is the elementary charge.

The matrix elements of the nuclear coupling Hamiltonian are expanded by
the eigenvalues and eigenvectors \cite{ccfull,Hagino12,Esb87}, and thus
are defined as
\begin{align}
 V^{(N)}_{nm}&=\langle n|V_N(r,\hat{O}_\lambda)|m\rangle-C_0(r) \notag\\
 &=\sum_\alpha\langle n|\alpha\rangle\langle\alpha|m\rangle V_N(r,\lambda_\alpha)-C_0(r), \label{cmat}
\end{align}
where $C_0$ is the coupling constant given by $C_0(r)=\langle
0|V_N(r,\hat{O}_\lambda)|0\rangle$.  In Ref.~\cite{Esb87}, the nuclear
coupling potential $V_N(r,\lambda_\alpha)=V_N^{(0)}(r-\lambda_\alpha)$
is expanded by $\lambda_\alpha$ and is taken up to the second order
of $\lambda_\alpha$, which is given by
\begin{equation}
 V_N^{}(r,\lambda_\alpha)= 
 V_N^{\rm (0)}(r)
 -\frac{dV_N^{\rm (0)}}{dr}\lambda_\alpha
+\frac{1}{2}\frac{d^2V_N^{\rm (0)}}{dr^2}\lambda_{\alpha}^{2}. \label{cpot}
\end{equation}
Thus, one can calculate $V_{nm}^{(N)}$ with Eqs.~(\ref{cmat}) and
(\ref{cpot}).  The Coulomb coupling matrix elements $V^{(C)}_{nm}$ are
similarly calculated using the linear coupling approximation
\cite{ccfull,Hagino12}.  The total coupling matrix elements are given by
the sum of $V_{nm}^{(N)}$ and $V_{nm}^{(C)}$.
 
Below, I describe an extension of the standard CC framework
following the strategy mentioned in the previous section. To this end, I
introduce the damping factor $\Phi(r,\lambda_\alpha)$ in the coupling
form factor.  Instead of Eq.~(\ref{cpot}), I employ the following form
for the nuclear coupling potential with respect to the eigen channel
$\alpha$,
\begin{equation}
 V_N^{}(r,\lambda_\alpha)= 
 V_N^{\rm (0)}(r)
 +\left[-\frac{dV_N^{\rm (0)}}{dr}\lambda_\alpha
+\frac{1}{2}\,\frac{d^2V_N^{\rm (0)}}{dr^2}\lambda_{\alpha}^{2}\right]\Phi(r,\lambda_\alpha).
\end{equation}
The most important modification to the standard CC treatment is the
introduction of the damping factor $\Phi$.  This damping factor
represents the physical process for gradually transitioning from sudden
to adiabatic approximations by diminishing the excitation strengths of
the target and/or projectile vibrational states after the two colliding
nuclei overlap one another.  To describe it, I choose the damping factor
as
\begin{equation}
 \Phi(r,\lambda_\alpha)=
  \begin{cases}
  e^{-(r-R_d-\lambda_\alpha)^2/2a_d^2}&\text{$r<R_d+\lambda_\alpha$} \\
    & \text{(Overlap region)}\\
  1 & \text{Otherwise} \\
    & \text{(Two-body region)},
   \label{dfact}
  \end{cases}           
\end{equation}
where $R_d$ is the spherical touching distance between the target and
the projectile defined by $R_d=r_d(A_{T}^{1/3}+A_{P}^{1/3})$. Here,
$r_d$ and $a_d$ are the damping radius and diffuseness parameters.  An
important point of these modifications is that the touching point in the
damping factor depends on $\lambda_\alpha$, namely, the excitation
strength begins to reduce at different distances in each eigen channel.

Therefore, in the two-body region ($r>R_d+\lambda_\alpha$), the standard
CC equations of Eq.~(\ref{cceq}) work well because $\Phi=1$.
Conversely, in the overlapping region ($r<R_d+\lambda_\alpha$), the
coupling matrix elements become $V_{nm}\rightarrow0$ because $\Phi
\rightarrow0$. Then, the standard CC equations of Eq.~(\ref{cceq}) are
close to the one-dimensional Schr\"odinger-like equations given by
\begin{multline}
\Biggl[-\frac{\hbar^2}{2\mu}\frac{d^2}{dr^2}
 +\frac{J(J+1)\hbar^2}{2\mu r^2}+V^{(0)}(r)
 +\epsilon_n-E\Biggr]u_n(r) = 0 \\
 (r\ll R_d+\lambda_\alpha). \label{cceq1bd}
\end{multline}
If an adiabatic one-body potential $V_{\rm 1bd}^{(0)}$ is substituted to
$V^{(0)}$ in Eq.~(\ref{cceq1bd}), one can avoid double counting of
the CC effects and correctly estimate the tunneling probability in the
one-body process. Subsequently, all the physical quantities are smoothly
joined.

It is technically complicated to take into account the effects of
damping factor on Coulomb coupling. I have introduced the channel
independent damping factor for the Coulomb coupling, but its effect on
fusion cross sections appear to be small. Therefore, I consider the
damping factor only for the nuclear coupling in the calculations
presented below.

Although I attempted to apply several functional forms to the damping
factor, I found that the form of Eq.~(\ref{dfact}) can well reproduce 
various experimental data.  Recently, the physical origin of the damping
factor was examined using the RPA method with a dinuclear shape
configuration \cite{ich13,ich15}. As shown in Eq.~(\ref{beta}), the
deformation parameter is directly related to the transition strength
$B(E\lambda)$. In Refs.~\cite{ich13,ich15}, the $B(E3)$ values for
individual colliding nuclei are directly calculated when they approach
each other. The obtained $B(E3)$ values drastically reduce near the
touching point and strongly correlate with the damping factor of
Eq.~(\ref{dfact}), which well reproduces the experimental fusion cross
sections for the $^{40}$Ca + $^{40}$Ca and $^{16}$O + $^{208}$Pb
reactions.  Namely, the damping factor describes the damping of quantum
vibrations near the touching point, indicating the suppression of
transitions between reaction channels in the CC equations. This
coordinate-dependent coupling strength would be responsible for the
fusion hindrance.

  \subsection{Heavy ion-ion potential}

In this paper, I adopt the Yukawa-plus-exponential (YPE)
potential~\cite{kra79} as a basic ion-ion potential $V_N^{(0)}$, because
the diagonal component of this potential satisfies conditions (1)-(3),
mentioned in Sec.~II-A by electing a suitable neck-formed shape for the
one-body system, as shown in our previous work~\cite{ich07-1}.  This
model is advantageous for a unified description of both two- and
one-body systems.  In this model, two Yukawa-type nuclear forces with a
different range parameter are assumed. One of the range parameters is
then determined by the saturation condition at the touching point.  In
Ref.~\cite{kra79}, the nuclear volume energy $E_{\rm V}$ is given by
\begin{equation}
 E_{\rm V}=-\frac{c_s}{8\pi^2r_0^2a^3}\int\int_V\left(\frac{\sigma}{a}-2\right)\frac{e^{-\sigma/a}}{\sigma}d^3rd^3r',
  \label{ypeint}
\end{equation}
where $\sigma=|\vec{r}-\vec{r'}|$, $r_0$ is the radius parameter, and
$a$ is the diffuseness parameter. In this paper, these two parameters
are adjustable. The nuclear radius is given by $R=r_0A^{1/3}$. The
integrations are performed over all nuclear densities.  The effective
surface constant $c_s$ is given by $c_s=a_s(1-\kappa_sI^2)$, where $a_s$
is the surface energy constant, $\kappa_s$ is the surface asymmetry
constant, and $I$ is the relative neutron-proton excess, and
$I=(N-Z)/A$. The values of $a_s$ and $\kappa_s$ are taken as $a_s=21.33$
MeV and $\kappa_s=2.378$ from the FRLDM2002 parameter set \cite{mo04}.

For two separated spherical nuclei of equivalent sharp surface radii
$R_T$ and $R_P$, the nuclear potential energy in the two-body system
$V_{\rm 2bd}^{(N)}$ before the touching point is given by
\begin{equation}
V_{\rm 2bd}^{(N)}(r) = -D\left(F+\frac{s}{a}\right)\frac{R_{TP}}{r}e^{-s/a},
\end{equation}
where $R_{TP}=R_T + R_P$, and $s=r-R_{TP}$.
The depth constant D is given by
\begin{equation}
D=\frac{4a^3g(R_T/a)g(R_P/a)e^{-R_{TP}/a}}{r_0^2R_{TP}}c_s',
\end{equation}
where $g(x)=x~{\rm cosh}(x)-{\rm sinh}(x)$ and $c_s'=\sqrt{c_s^{T}c_s^{P}}$.
The constant $F$ is given by
\begin{equation}
 F=4+\frac{R_{TP}}{a}-\frac{f(R_T/a)}{g(R_T/a)}-\frac{f(R_P/a)}{g(R_P/a)},
\end{equation}
where $f(x)=x^2~{\rm sinh}(x)$.

For the one-body potential, one can calculate $V_{\rm 1bd}(N)$ by
integrating Eq.~(\ref{ypeint}) with an appropriately shaped
parametrization having a neck formation, such as in the previous work of
Ref.~\cite{ich07-1}.  However, this calculation is time-consuming for
systematic investigations.  For mass-asymmetric systems, it is also
difficult to smoothly joint the potential energies between the two-body
and the adiabatic one-body systems at the touching point, because the
proton-to-neutron ratio for the one-body system differs from that for
the target and projectile in the two-body system.  There is
discontinuity even at the touching point in symmetrical systems for a few
physical quantities, including the Wigner term and the $A_0$ constant in
the nuclear mass model (for details see Ref.~\cite{mo04}).

To avoid these difficulties, I approximate the one-body potential with a
third-order polynomial function based on the YPE potential using the
lemniscatoids parametrization~\cite{Roy82,Roy84,Roy85} (see Appendices A
and B for detail).  I smoothly joint the potential energy at the
touching point to the energy of its compound state, estimated from
experimental nuclear masses. I also perform this by identifying the
internucleus distance $r$ with the centers-of-masses distance of two
half spheres.  I have systematically tested the performance of this
procedure for various systems.  The deviation due to this procedure is
negligible.  This procedure works very well in many systems, because the
data points at the lowest incident energy in the experiments which have
been performed until now are less than the potential energies at the
touching point only by a few MeV. At much deeper incident energies, the
adiabatic one-body potential energy would play a decisive role in the
fusion hindrance.

In this procedure, there are three input parameters: the position of the
compound state $r_{\rm GS}$, the energy of the compound state $E_{\rm
GS}$, and the potential energy curvature at the ground state
$\hbar\omega_{\rm GS}$.  The value of $r_{\rm GS}$ is estimated by the
center-of-mass distance between the two halves of its spherical compound
nucleus, which is given by $r_{\rm GS}=3/4\cdot R_c$, where
$R_c=r_0(A_T+A_P)^{1/3}$.  The value of $E_{\rm GS}$ is estimated from
the experimental nuclear masses taken from the AME2003 table
\cite{audi03}.  If the experimental mass value is not available for a
nucleus, I use the calculated mass from the FRDM95 table \cite{Mol95}.
The value of $\hbar\omega_{\rm GS}$ is estimated from a systematic curve
fitted to the curvatures of the liquid drop energies at their ground
states for various systems (see Appendix B).  This is given by
$\hbar\omega_{\rm GS}=0.0047x^2-0.4586x+9.125$ MeV, where
$x=A_T^{1/3}\cdot A_P^{1/3}$.

\begin{figure}[htbp]
 \includegraphics[keepaspectratio,width=7cm]{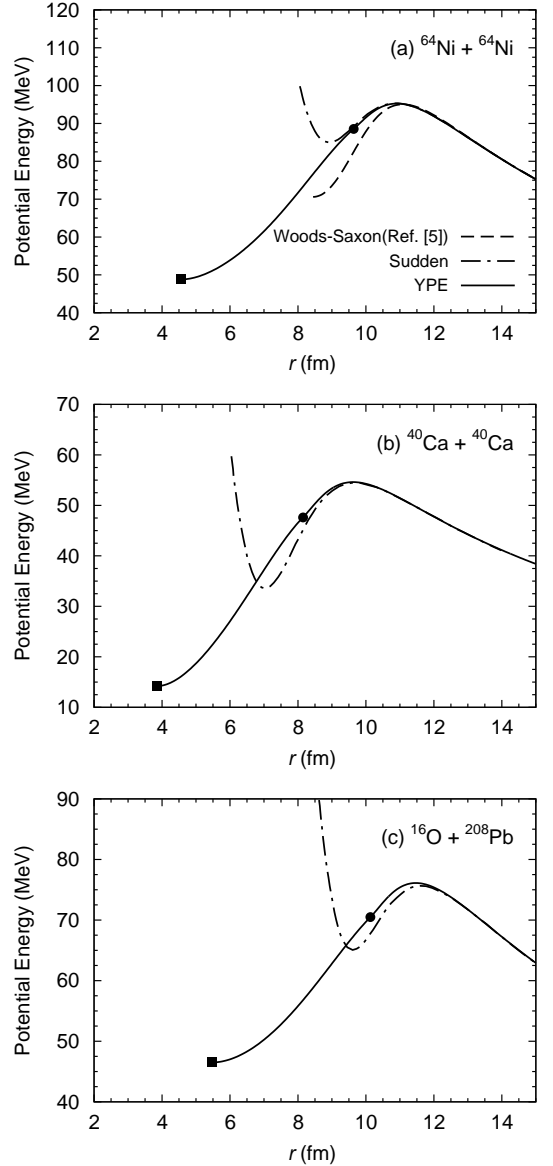} \caption{
 Potential energies versus the center-of-mass distance $r$ for the (a)
 $^{64}$Ni + $^{64}$Ni, (b) $^{40}$Ca + $^{40}$Ca, and (c) $^{16}$O +
 $^{208}$Pb systems.  The solid lines indicate the calculated YPE
 potential. The solid circles and squares denote the potential energy
 at the touching point and the compound state estimated from
 experimental nuclear masses, respectively.  } \label{pot}
\end{figure}

Figure \ref{pot} shows the calculated YPE potentials (solid lines) for
the (a) $^{64}$Ni + $^{64}$Ni, (b) $^{40}$Ca + $^{40}$Ca, and (c)
$^{16}$O + $^{208}$Pb systems.  In the calculations, I use the $r_0$ and
$a$ fitted parameters to reproduce their experimental data by CC
calculations (this is discussed later, see the parameters in the tables
in Sec. III).  From the figure, it is evident that the calculated
potential energies in the two-body system are smoothly jointed to its
compound states (solid squares) at the touching point (solid circles).
For comparison, the potential energies calculated with the Woods-Saxon
(WS) and sudden models are demonstrated by the dashed and dash-dotted
lines, respectively.  The parameters of the WS model are taken from
Ref.~\cite{Jiang04}.  The results calculated with the sudden model are
from Refs.~\cite{mis06,Esb07,Monta12}.  In Fig.~\ref{pot} (a), the YPE
potential around the Coulomb barrier is recognizably thicker than that
of the WS model using the parameters from Ref.~\cite{Jiang04}.

Interestingly, the YPE potentials are similar to the sudden ones
before the touching point.  In Fig.~\ref{pot} (a), the YPE potential is
almost identical to the sudden one before the touching point.  In
Figs.~\ref{pot} (b) and (c), the YPE potentials are also similar to the
sudden ones before the touching point.  After the touching point,
the sudden potentials have a shallow potential pocket, whereas the
adiabatic potentials have different energy dependence and a much
deeper potential pocket.  This large difference at smaller $r$ indicates
that the mechanisms for the fusion hindrance in the two models are
completely different.

 \section{Calculation Results}

I perform the CC calculations with the damping factor described in the
previous section for the medium-heavy mass systems of $^{64}$Ni +
$^{64}$Ni, $^{58}$Ni + $^{58}$Ni, and $^{58}$Ni + $^{54}$Fe: the
medium-light mass systems of $^{40}$Ca + $^{40}$Ca, $^{48}$Ca +
$^{48}$Ca, and $^{24}$Mg + $^{30}$Si: and the mass-asymmetric systems of
$^{48}$Ca + $^{96}$Zr and $^{16}$O + $^{208}$Pb.

For this purpose, I implemented the YPE potential and the damping factor in
the computer code {\tt CCFULLYPE} \cite{ccfullype}, which is a modified
version of the code {\tt CCFULL} \cite{ccfull}.  Note that the
definition for the origin of the coupling potential ($C_0$ in
Eq.~\ref{cmat}) is different from that of the original {\tt CCFULL} in
Ref.~\cite{ccfull}. To compare my calculated results with those of the
sudden model, I adopt the definition used in the sudden model
\cite{mis06}.

I calculate the fusion cross sections for these systems and show the
astrophysical S factor and logarithmic derivative representations of the
obtained fusion cross sections.  In this paper, I focus the discussion
on fusion cross sections from the sub-barrier to deep sub-barrier
incident energies, because the adiabatic approach can work well in this
energy region. At incident energies much higher than the Coulomb
barrier, where oscillations of the fusion excitation function have been
recently studied \cite{Monta15}, other approximations, including the
sudden model, would be more appropriate.

In this paper, the S factor is given by $S(E)=E\sigma_{\rm
fus}(E)\exp{(2\pi(\eta-\eta_0))}$, where $E$ is the incident energy,
$\sigma_{\rm fus}$ is the fusion cross section, $\eta$ is the Sommerfeld
parameter, and $\eta_0$ is an arbitrary unit.  The Sommerfeld parameter
is given by $\eta=e^2Z_tZ_p/\hbar\nu$, where $Z_t$ and $Z_p$ are the
charges of the target and projectile, respectively, and $\nu$ is the
relative velocity between the target and projectile in the
center-of-mass frame.  The logarithmic derivative of the fusion cross
section is given by $L(E)=\frac{d}{dE}\ln(E\sigma_{\rm fus}(E))$.

\subsection{Medium-heavy mass systems}

\begin{table}
 \caption{\label{cc1} Input parameters for the coupling strengths in the
 CC calculations for the $^{64}$Ni + $^{64}$Ni, $^{58}$Ni + $^{58}$Ni,
 and $^{58}$Ni + $^{54}$Fe systems. The symbol $\lambda^\pi$ denotes the
 multipolarity and the parity of a state. The symbol $E_{\rm ex}$
 denotes the excitation energy of a state. The symbols
 $\beta_{\lambda}^{\rm Coul}$ and $\beta_{\lambda}^{\rm Nucl}$ denote
 the deformation parameters for the Coulomb and nuclear coupling
 strengths, respectively. The symbol $N_{\rm ph}$ denotes the number of
 phonons included in the calculations.}
 \begin{ruledtabular}
   \begin{tabular}{cccccc}
    Nucleus&$\lambda^\pi$&$E_{\rm ex}$ (MeV)& $\beta_{\lambda}^{\rm Coul}$ &
    $\beta_{\lambda}^{\rm Nucl}$ & $N_{\rm ph}$\\ \hline
    \multicolumn{6}{l}{(a) $^{64}$Ni + $^{64}$Ni (see Ref. \cite{Jiang04})}\\
     $^{64}$Ni & $2^+$ & 1.346 & 0.165 & 0.185 & 2 \\
          &$3^-$ & 3.560 & 0.193 & 0.200 & 1 \\
    \multicolumn{5}{l}{(b) $^{58}$Ni + $^{58}$Ni (see
    Ref. \cite{Esb87})} \\ 
     $^{58}$Ni & $2^+$ & 1.450 & 0.187 & 0.226 & 3 \\
     & $3^-$ & 4.470 & 0.200 & 0.200 & 1 \\
    \multicolumn{5}{l}{(c) $^{58}$Ni + $^{54}$Fe (see
    Refs.~\cite{Esb87,Stef10})} \\ 
    $^{58}$Ni & $2^+$ & 1.450 & 0.187 & 0.187 & 2 \\
    & $3^-$ & 4.470 & 0.200 & 0.200 & 1 \\
    $^{54}$Fe & $2^+$ & 1.408 & 0.200 & 0.200 & 1 \\
   \end{tabular}
 \end{ruledtabular}
\end{table}

\begin{table}
 \caption{\label{ype1} Input parameters for the YPE potential, the
 damping factor in the $^{64}$Ni + $^{64}$Ni, $^{58}$Ni + $^{58}$Ni, and
 $^{58}$Ni + $^{54}$Fe systems. The symbols $r_0$ and $a_0$ denote the
 radius and diffuseness parameters in the YPE potential. The symbols
 $V_{\rm GS}$ and $\hbar\omega_{\rm GS}$ denote the potential energy and
 its curvature at each ground state. The symbol $V_{\rm Touch}$ denotes
 the potential energy at the touching point. The symbols $r_d$ and $a_d$
 denote the radius and diffuseness parameters in the damping factor.}
 \begin{ruledtabular}
    \begin{tabular}{ccccccccc}
     System & $r_0$ & $a_0$ & $V_{\rm GS}$& $\hbar\omega_{\rm GS}$
     &$V_{\rm Touch}$ & $r_d$ & $a_d$ \\
     & (fm) & (fm) & (MeV) & (MeV) &(MeV)& (fm) & (fm) \\\hline
    $^{64}$Ni + $^{64}$Ni & 1.205 &0.68&48.8&2.99&88.6&1.298&1.05\\
    $^{58}$Ni + $^{58}$Ni & 1.180 &0.68&66.1&3.31&95.3&1.310&1.32\\
    $^{58}$Ni + $^{54}$Fe & 1.198 &0.68&58.7&3.42&86.8&1.330&1.25\\
    \end{tabular}
 \end{ruledtabular}
\end{table}

First, I discuss the fusion hindrance in the medium-heavy mass systems
for the $^{64}$Ni + $^{64}$Ni, $^{58}$Ni + $^{58}$Ni, and $^{58}$Ni +
$^{54}$Fe systems.  All input parameters for the CC calculations are
tabulated in Tables.~\ref{cc1} and \ref{ype1}.  In the calculations, I
included couplings only to the low-lying $2^+$ and $3^-$ states and all
mutual excitations of these states. The deformation parameters are
basically the same as those in Refs.~\cite{Jiang04,Esb87,Stef10}. Only
for the $2^+$ state of $^{58}$Ni in the $^{58}$Ni + $^{58}$Ni system, I
use a 20\% larger value of $\beta_{\lambda}^{\rm Nucl}$ compared to
$\beta_{\lambda}^{\rm Coul}$. 

\begin{figure}[htbp]
 \includegraphics[keepaspectratio,width=7cm]{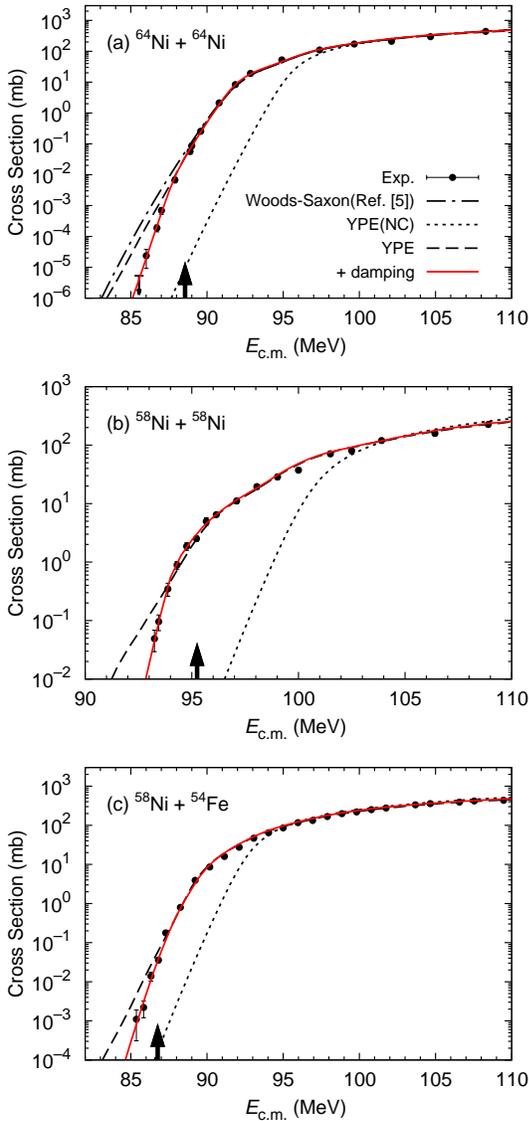}
 \caption{(Color online) Calculated fusion cross sections versus
 incident energies for the (a) $^{64}$Ni + $^{64}$Ni, (b) $^{58}$Ni +
 $^{58}$Ni, and (c) $^{58}$Ni + $^{54}$Fe systems. The solid circles
 denote the experimental data. The solid and dashed lines indicate the
 calculated results with and without the damping factor,
 respectively. The dotted lines indicate the calculated results of no
 coupling. The dash-dotted line indicates the result calculated with
 the WS potential. The arrows indicate the potential energy at the
 touching point $V_{\rm Touch}$.}  \label{cross1}
\end{figure}

Figure \ref{cross1} shows the calculated fusion cross sections.  The
solid and dashed lines indicate the results calculated with and without
the damping factor, respectively.  The dotted lines indicate the results
calculated with no coupling.  The experimental data for $^{64}$Ni +
$^{64}$Ni, $^{58}$Ni + $^{58}$Ni, and $^{58}$Ni + $^{54}$Fe (solid
circles) are taken from Refs.~\cite{Jiang04,Beck81,Beck82,Stef10},
respectively.  In the figure, one can see that drastic improvements have
been made by taking into account the damping in the CC form factor, as
compared to the results calculated without the damping factor.  I tested
the dependence of $\hbar\omega_{\rm GS}$ on the calculated results, but
it was negligible above the lowest incident energy of the experimental
data in each system.  Interestingly, in the $^{58}$Ni + $^{58}$Ni
system, the fusion cross sections are slightly enhanced around an
incident energy of 95 MeV due to the damping factor.

For comparison, in Fig.~\ref{cc1} (a), I also plot the result calculated
with the WS potential using the parameters given in Ref.~\cite{Jiang04}.
Since the YPE potential is much thicker than the WS potential, as shown
in Fig.~\ref{pot}, the result calculated with the YPE potential is
slightly suppressed. Although the potential thickness around the Coulomb
barrier increases considerably in the YPE potential, one cannot
reproduce the fusion cross sections at the deep sub-barrier energies
only by changing it. The damping factor plays an important role in
reproducing the fusion hindrance behavior.

In each of these systems, $V_{\rm Touch}$ remarkably correlates with the
incident threshold energy for fusion hindrance. The values of
$V_{\rm Touch}$ are tabulated in Tabel~\ref{ype1} and indicated by the
arrows in Fig.~\ref{cross1}.  In all the systems, one can clearly see
that the significant decreases in the fusion cross sections start from
$V_{\rm Touch}$. Thus, the threshold rule for the potential energy at
the touching point works very well in the medium-heavy mass systems.

\begin{figure*}[htbp]
 \includegraphics[keepaspectratio,width=13cm]{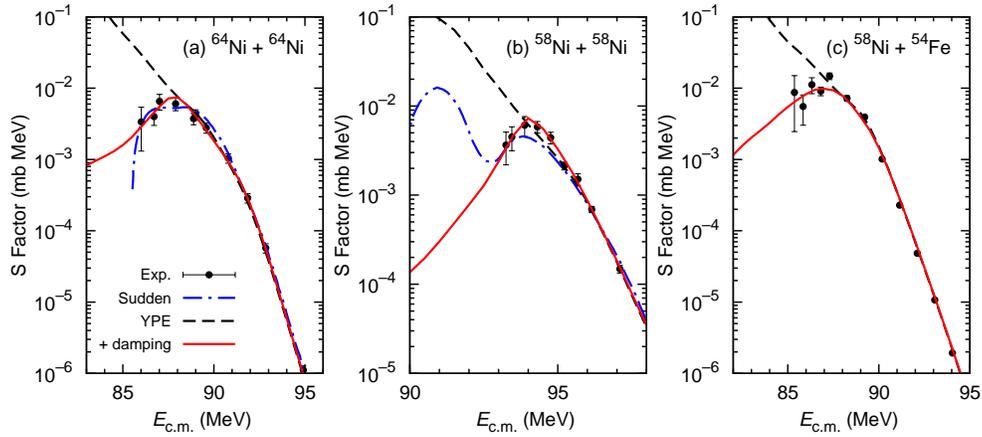} \caption{(Color
 online) Astrophysical S factor representations of the fusion cross
 sections versus incident energies for the (a) $^{64}$Ni + $^{64}$Ni,
 (b) $^{58}$Ni + $^{58}$Ni, and (c) $^{58}$Ni + $^{54}$Fe systems. The
 solid and dashed lines indicate the results calculated with and
 without the damping factor, respectively. The dash-dotted lines
 indicate the results calculated with the sudden model from
 Refs.~\cite{mis06}} \label{sfact1}
\end{figure*}

Figure \ref{sfact1} shows the astrophysical S factor representations of
the fusion cross sections for these systems. I take $\eta_0=$ 75.23,
69.99, and 66.0 MeV for the $^{64}$Ni + $^{64}$Ni, $^{58}$Ni +
$^{58}$Ni, and $^{58}$Ni + $^{54}$Fe systems, respectively. In the
figure, the solid and dashed lines indicate the results calculated with
and without the damping factor, respectively.  In all the systems, the
results calculated with the damping factor are in good agreement with
the experimental data.  In each of these systems, the calculated result
well reproduces the single peak structure of the experimental data.  For
comparison, the results calculated with the sudden model from
Ref.~\cite{mis06} are plotted by the dash-dotted lines in
Figs.~\ref{sfact1} (a) and (b).  In both the systems, the adiabatic
model better reproduces the experimental data compared to the sudden
model.  The S factors of the adiabatic and sudden models at the deep
sub-barrier energies are considerably different from each other,
although the reproductions of the fusion cross sections calculated with
the two models are similar to an extent.

In the $^{64}$Ni + $^{64}$Ni system, the S factor calculated with the
sudden model significantly decreases with decreasing incident energy,
whereas the adiabatic model has a much weaker and smoother energy
dependence. In the $^{58}$Ni + $^{58}$Ni system, unphysical fluctuations
of the S factor are recognizable at extremely low incident energies in
the sudden model, whereas the adiabatic model has a single, smooth peak.
Thus, fusion cross section measurements at much deeper incident energies
for this system are appropriate for discriminating which model can
better describe the deep sub-barrier fusions.

\begin{figure*}[htbp]
 \includegraphics[keepaspectratio,width=13cm]{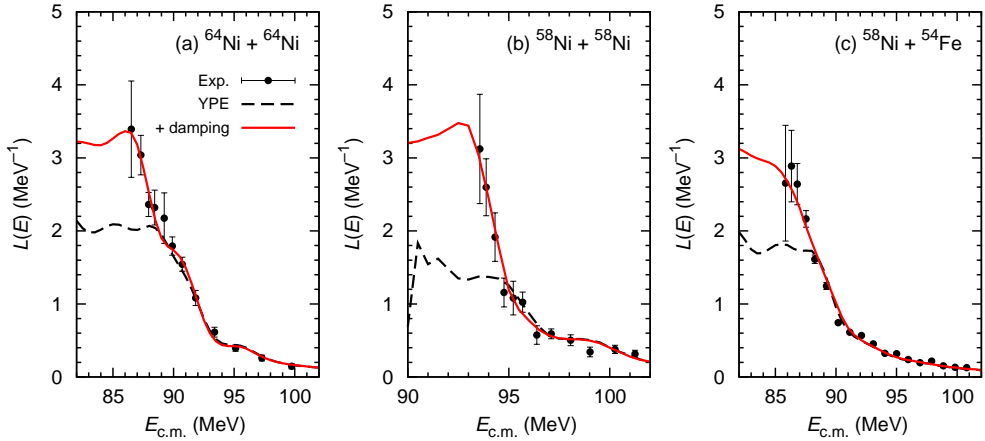} \caption{(Color
 online) Logarithmic derivative representations of the fusion cross
 sections versus incident energies for the (a) $^{64}$Ni + $^{64}$Ni,
 (b) $^{58}$Ni + $^{58}$Ni, and (c) $^{58}$Ni + $^{54}$Fe systems. The
 solid and dashed lines indicate the results calculated with and
 without the damping factor, respectively.}  \label{log1}
\end{figure*}

Figure \ref{log1} shows the logarithmic derivative representations of
the fusion cross sections versus incident energies.  The solid and
dashed lines indicate the results calculated with and without the
damping factor, respectively. All the calculated results are in good
agreement with the experimental data. In each of these systems, the
result calculated with the damping factor is saturated below a certain
incident energy. Conversely, the result calculated with the sudden
model significantly increases or fluctuates with decreasing incident
energy~\cite{mis06}.  At the deep sub-barrier energies, the energy
dependence of the logarithmic derivative in the adiabatic model is
substantially different from that in the sudden model.
 
\subsection{Medium-light mass systems}

\begin{table}
 \caption{\label{cc2} Input parameters for the coupling strengths in the
 CC calculations for the $^{40}$Ca + $^{40}$Ca, $^{48}$Ca + $^{48}$Ca,
 and $^{24}$Mg + $^{30}$Si systems. All symbols are the same as those in
 Table~\ref{cc1}.}
 \begin{ruledtabular}
   \begin{tabular}{cccccc}
    Nucleus&$\lambda^\pi$&$E_{\rm ex}$ (MeV)& $\beta_{\lambda}^{\rm Coul}$ &
    $\beta_{\lambda}^{\rm Nucl}$ & $N_{\rm ph}$\\ \hline
    \multicolumn{6}{l}{(a) $^{40}$Ca + $^{40}$Ca (see
    Refs.~\cite{Monta12, Esb89})}\\
     $^{40}$Ca & $2^+$ & 3.905 & 0.119 & 0.119 & 1 \\
    &$3^-$ & 3.737 & 0.402 & 0.402 & 1 \\
    \multicolumn{5}{l}{(b) $^{48}$Ca + $^{48}$Ca (see
    Ref. \cite{Esb10})} \\ 
     $^{48}$Ca & $2^+$ & 3.832 & 0.102 & 0.154 & 2 \\
     & $3^-$ & 4.507 & 0.203 & 0.154 & 1 \\
    \multicolumn{5}{l}{(c) $^{24}$Mg + $^{30}$Si (see
    Ref. \cite{Jiang14})} \\ 
    $^{24}$Mg & $2^+$ & 1.369 & 0.608 & 0.460 & 1 \\
    $^{30}$Si & $2^+$ & 2.235 & 0.330 & 0.330 & 1 \\
    & $3^-$ & 5.497 & 0.275 & 0.275 & 1 \\
   \end{tabular}
 \end{ruledtabular}
\end{table}

\begin{table}
 \caption{\label{ype2} Input parameters for the YPE potential and the
 damping factor in the $^{40}$Ca + $^{40}$Ca, $^{48}$Ca + $^{48}$Ca, and
 $^{24}$Mg + $^{30}$Si systems. All symbols are the same as those in 
 Tabel~\ref{ype1}.}
 \begin{ruledtabular}
    \begin{tabular}{ccccccccc}
     System & $r_0$ & $a_0$ & $V_{\rm GS}$& $\hbar\omega_{\rm GS}$
     &$V_{\rm Touch}$ & $r_d$ & $a_d$ \\
     & (fm) & (fm) & (MeV) & (MeV) &(MeV)& (fm) & (fm) \\\hline
    $^{40}$Ca + $^{40}$Ca & 1.191 &0.68&14.2&4.40 &47.6 &1.240&0.52\\
    $^{48}$Ca + $^{48}$Ca & 1.185 &0.68&3.0 &3.89 &43.0 &1.280&0.60\\
    $^{24}$Mg + $^{30}$Si & 1.190 &0.68&$-17.9$&5.39 &15.8&1.430&1.25\\
    \end{tabular}
 \end{ruledtabular}
\end{table}

Next, I discuss the fusion hindrance in the medium-light mass systems
for the $^{40}$Ca + $^{40}$Ca, $^{48}$Ca + $^{48}$Ca, and $^{24}$Mg +
$^{30}$Si systems. These systems are appropriate to check the dependence
of the reaction Q value on the fusion hindrance. The reaction Q values
of the $^{40}$Ca + $^{40}$Ca, $^{48}$Ca + $^{48}$Ca, and $^{24}$Mg +
$^{30}$Si systems are $Q=-14.2$, $-3.0$, and 17.9 MeV, respectively.
All input parameters for the coupling strengths of the CC calculations
are tabulated in Table~\ref{cc2}.  In the calculations, I included
couplings only to the low-lying $2^+$ and $3^-$ states and all mutual
excitations of these states.  For $^{40}$Ca + $^{40}$Ca, I used the same
deformation parameters between $\beta^{\rm Coul}$ and $\beta^{\rm
Nucl}$, which differ from those of Ref.~\cite{Monta12}.  All input
parameters for the YPE potential and the damping factor are tabulated in
Table~\ref{ype2}.

\begin{figure}[htbp]
 \includegraphics[keepaspectratio,width=7cm]{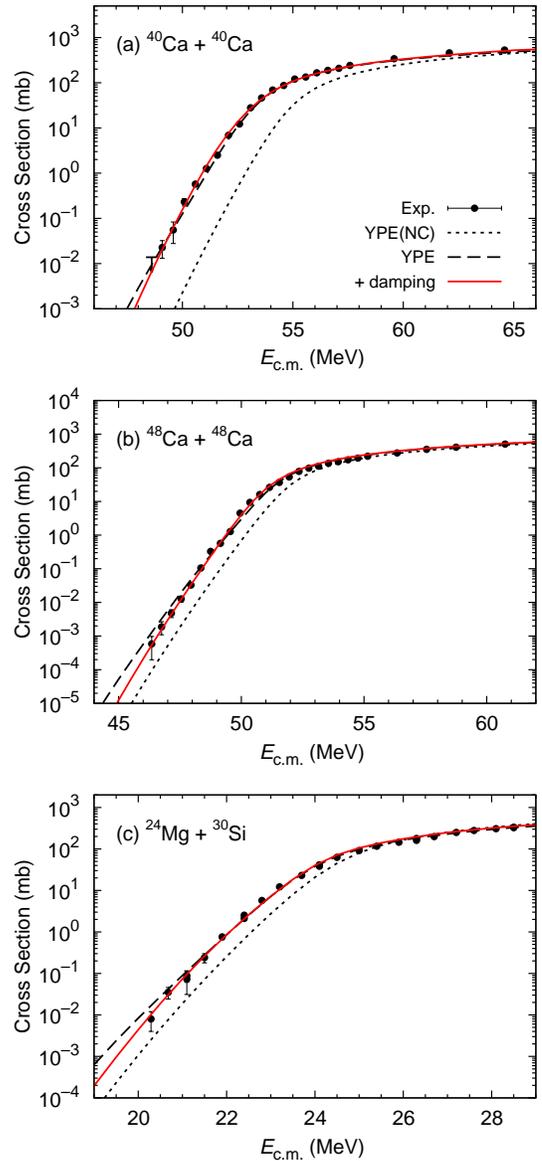} \caption{(Color
 online) Calculated fusion cross section versus incident energies for
 the (a) $^{40}$Ca + $^{40}$Ca, (b) $^{48}$Ca + $^{48}$Ca, and (c)
 $^{24}$Mg + $^{30}$Si systems.  All symbols are the same as those in
 Fig.~\ref{cross1}.}  \label{cross2}
\end{figure}

Figure \ref{cross2} shows the obtained fusion cross sections. All
the calculated results (solid lines) are in good agreement with the
experimental data (solid circles).  The experimental data for the
$^{40}$Ca + $^{40}$Ca, $^{48}$Ca + $^{48}$Ca, and $^{24}$Mg + $^{30}$Si
systems are from Refs.~\cite{Monta12}, Ref.~\cite{Stef09}, and
Ref.~\cite{Jiang14}, respectively.  In comparison to the medium-heavy
mass systems, the effect of the damping factor on the fusion cross
sections is relatively small in each system (dashed lines in
Fig.~\ref{cross2}). This is because that the potential energies at the
touching point are lower than the lowest incident energies in the
available experimental data ($V_{\rm Touch}$ in Table~\ref{ype2}).
Thus, the calculated results presented in the energy regions of
Fig.~\ref{cross2} are independent of the slope of the one-body potential
around the touching point associated with $\hbar\omega_{\rm GS}$.  In
this regard, it seems that the fusion hindrance is relatively weak in
these systems.

The potential energies at the touching point in these system weakly
correlate with the threshold incident energies for the fusion
hindrance. In each of these systems, the value of $V_{\rm Touch}$ is
lower than the threshold incident energy for fusion hindrance
by about 3$\sim$5 MeV.  In these systems, the deformation parameters in
the coupling strengths from the CC calculations are considerably
large. This implies the effect of the damping factor starts much before
the touching point.  Thus, the adiabatic potential in the CC model is
affected by these large deformation effects, which is discussed later in
Sec.~IV-A.  In these medium-light mass systems, the threshold rule
should be modified by taking into account the deformation effects.
 
\begin{figure*}[htbp]
 \includegraphics[keepaspectratio,width=13cm]{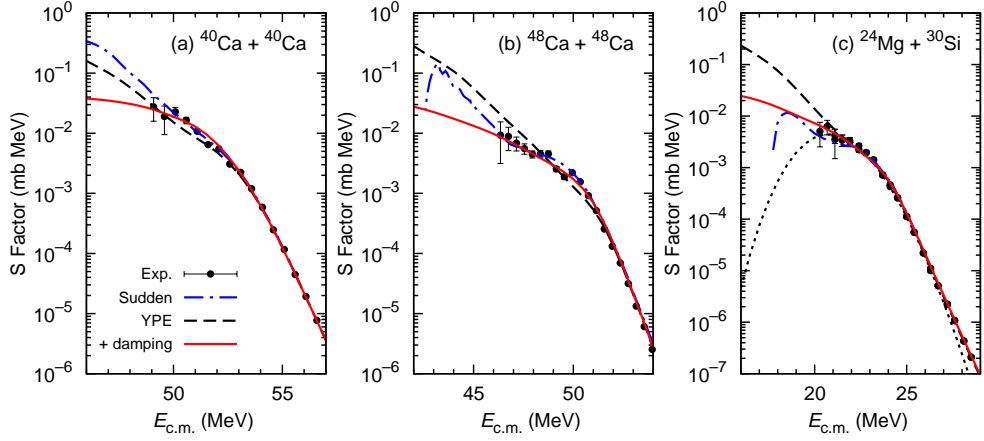} \caption{(Color
 online) Astrophysical S factor representations of the fusion cross
 sections versus incident energies for the (a) $^{40}$Ca + $^{40}$Ca,
 (b) $^{48}$Ca + $^{48}$Ca, and (c) $^{24}$Mg + $^{30}$Si systems. All
 symbols are the same as those in Fig.~\ref{sfact1}. The dotted line
 indicates the functional form proposed by Jiang {et
 al.}~\cite{Jiang14}.}\label{sfact2}
\end{figure*}

Figure \ref{sfact2} shows the astrophysical S factor representations of
the fusion cross sections versus incident energies for these systems. I
take $\eta_0= 40.8$, 45.5, and 22.0 for the $^{40}$Ca + $^{40}$Ca,
$^{48}$Ca + $^{48}$Ca, and $^{24}$Mg + $^{30}$Si systems, respectively.
All the calculated results (solid lines) are in good agreement with
the experimental data (solid circles). For comparison, the S factors
calculated with the sudden model taken from
Refs.~\cite{Monta12,Esb10,Jiang14} are plotted by the dash-dotted
lines. The S factors calculated with the damping factor are in good
agreement with the experimental data compared to those with the sudden
model.  As discussed in the medium-heavy mass systems, the S factors
calculated with the damping factor have a much smoother
energy dependence than those with the sudden model.

In each of these systems, the peak structure in the S factor calculated
with the damping factor is not visible, although Jiang {\it et al.}
assumed it was in their fitting function~\cite{Jiang14}.  For
comparison, the S factor proposed by Jiang {\it et al.} is plotted by
the dotted line in Fig.~\ref{sfact2} (c). The S factor of Jiang {\it et
al.} has a strong energy dependence and peak structure. However, there
is no physical reason why the S factor can be described by the
functional form proposed by them.  This large difference in S factor
behavior at the low incident-energy region strongly affects the
estimations of astrophysical reaction rates.  Nevertheless, a few
additional data points with high precision at the low-energy region are
necessary to determine this behavior.

\begin{figure*}[htbp]
 \includegraphics[keepaspectratio,width=13cm]{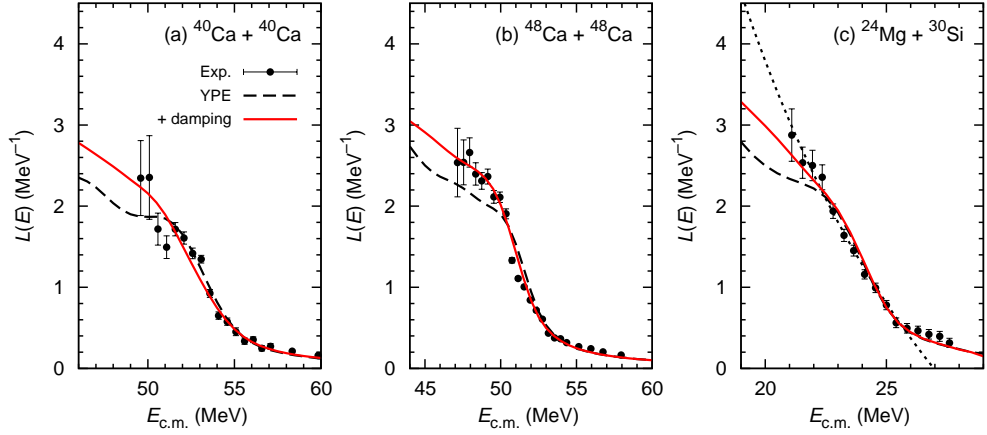} \caption{(Color
 online) Logarithmic derivative representations of the fusion cross
 sections versus incident energies for the (a) $^{40}$Ca + $^{40}$Ca,
 (b) $^{48}$Ca + $^{48}$Ca, and (c) $^{24}$Mg + $^{30}$Si systems. All
 symbols are the same as those in Fig.~\ref{log1}.}  \label{log2}
\end{figure*}

Figure \ref{log2} shows the logarithmic derivative representations of
the calculated fusion cross sections for the $^{40}$Ca + $^{40}$Ca,
$^{48}$Ca + $^{48}$Ca, and $^{24}$Mg + $^{30}$Si systems. The results
calculated with the damping factor (solid lines) are in good agreement
with the experimental data (solid circles).  For comparison, the
logarithmic derivative presented by Jiang {\it et al.}~is also plotted
by the dotted line.  In each of these systems, the calculated S factor
monotonically increases with decreasing incident energy and the slope
changes at a certain energy.  In contrast, the logarithmic derivative of
Jiang {\it et al.}~linearly increases with decreasing incident energy.

In the adiabatic model presented in this paper, the reaction Q value is
an important input parameter in the heavy ion-ion potential for
estimating the ground-state energy of the compound system.  In these
systems, the compound state energies are sufficiently lower than the
potential energies at the touching point (Table~\ref{ype2}). Therefore,
the dependence of the Q values on the fusion cross sections is
negligible in the adiabatic model, although Jiang discussed the effect
of the positive Q value on the fusion hindrance~\cite{Jiang14}.

\subsection{Mass-asymmetric reaction systems}

\begin{table}
 \caption{\label{cc3} Input parameters for the coupling strengths in the
 CC calculations for the $^{48}$Ca + $^{96}$Zr and $^{16}$O + $^{208}$Pb
 systems. All symbols are the same as those in Table.~\ref{cc1}.}
 
 \begin{ruledtabular}
   \begin{tabular}{cccccc}
    Nucleus&$\lambda^\pi$&$E_{\rm ex}$ (MeV)& $\beta_{\lambda}^{\rm Coul}$ &
    $\beta_{\lambda}^{\rm Nucl}$ & $N_{\rm ph}$\\ \hline
    \multicolumn{6}{l}{(a) $^{48}$Ca + $^{96}$Zr (see Refs.~\cite{Esb09})}\\
     $^{48}$Ca   &$2^+$ & 3.832 & 0.102 & 0.126 & 2 \\
     $^{96}$Zr & $2^+$ & 1.751 & 0.079 & 0.079 & 1 \\
               & $3^-$ & 1.897 & 0.295 & 0.295 & 3 \\
    \multicolumn{6}{l}{(b) $^{16}$O + $^{208}$Pb (see Refs.~\cite{Esb07})}\\
     $^{16}$O   &$3^-$ & 6.129 & 0.713 & 0.713 & 2 \\
     $^{208}$Pb & $3^-$ & 2.615 & 0.111 & 0.111 & 2 \\
   \end{tabular}
 \end{ruledtabular}
\end{table}

\begin{table}
 \caption{\label{ype3} Input parameters for the YPE potential and the
 damping factor in the $^{48}$Ca + $^{96}$Zr and $^{16}$O + $^{208}$Pb
 systems. All symbols are the same as those in Table.~\ref{ype1}}
 \begin{ruledtabular}
    \begin{tabular}{ccccccccc}
     System & $r_0$ & $a_0$ & $V_{\rm GS}$& $\hbar\omega_{\rm GS}$
     &$V_{\rm Touch}$ & $r_d$ & $a_d$ \\
     & (fm) & (fm) & (MeV) & (MeV) &(MeV)& (fm) & (fm) \\\hline
     $^{48}$Ca + $^{96}$Zr & 1.198 &0.68&45.9&3.16 &88.8 &1.30&1.05\\
     $^{16}$O + $^{208}$Pb & 1.20 &0.68&46.5&3.33 &70.5 &1.255&1.14\\
    \end{tabular}
 \end{ruledtabular}
\end{table}

Finally, I discuss the mass-asymmetric reaction system for the $^{48}$Ca
+ $^{96}$Zr and $^{16}$O + $^{208}$Pb systems.  All input parameters for
the coupling strengths in the CC calculations are tabulated in
Table.~\ref{cc3}. In the calculations, I included couplings only to the
low-lying $2+$ and $3^-$ states and all mutual excitations of these
states.  For $^{16}$O + $^{208}$Pb, I included only the $3^-$ states of
$^{16}$O and $^{208}$Pb and used the same deformation parameters of
$\beta^{\rm Coul}$ and $\beta^{\rm Nucl}$, although those of
Ref.~\cite{Esb07} are different.

\begin{figure}[htbp]
 \includegraphics[keepaspectratio,width=7cm]{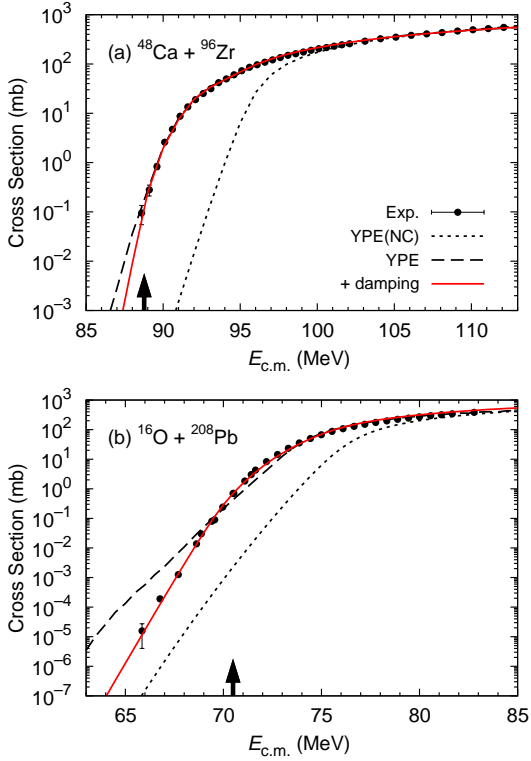}
 \caption{(Color online) Calculated fusion cross sections versus
 incident energies for the (a) $^{48}$Ca + $^{96}$Zr and (b) $^{16}$O +
 $^{208}$Pb systems. All symbols are the same as those in
 Fig.~\ref{cross1}.} \label{cross3}
\end{figure}

Figure \ref{cross3} shows the calculated fusion cross sections versus
incident energies. The results calculated with the damping factor (solid
lines) are in good agreement with the experimental data (solid circles).
The experimental data for $^{48}$Ca + $^{96}$Zr and $^{16}$O +
$^{208}$Pb is from Ref.~\cite{Stef06} and Refs.~\cite{das07,Morton99},
respectively.  In the $^{16}$O + $^{208}$Pb system, it is seen that a
drastic improvement has been made by taking into account the damping
factor in the CC form factor, as compared to the result without the
damping factor (dashed line).  A strong fusion hindrance can be seen in
the $^{16}$O + $^{208}$Pb system.
 
In these systems, the potential energies at the touching point strongly
correlate with the threshold incident energies for fusion hindrance.
The values of $V_{\rm Touch}$ are tabulated in Table.~\ref{ype3} and are
indicated by the solid arrows in Fig.~\ref{cross3}.  The threshold rule
works well in these systems.  The experimental data for the $^{16}$O +
$^{208}$Pb system are most adequate for discussing the fusion hindrance,
because the lowest incident energy in the experimental data is lower
than $V_{\rm Touch}$ by about 5 MeV. Only this measurement has achieved
such deep sub-barrier energy.  This is because that the position of
$V_{\rm Touch}$ approaches that of the Coulomb barrier as the mass
number of the compound system increases. Thus, the fusion hindrance
would be more clearly observed in such heavy-mass compound system. The
fusion hindrance phenomena would play a decisive role in the formation
of super heavy elements.

\begin{figure}[htbp]
 \includegraphics[keepaspectratio,width=\linewidth]{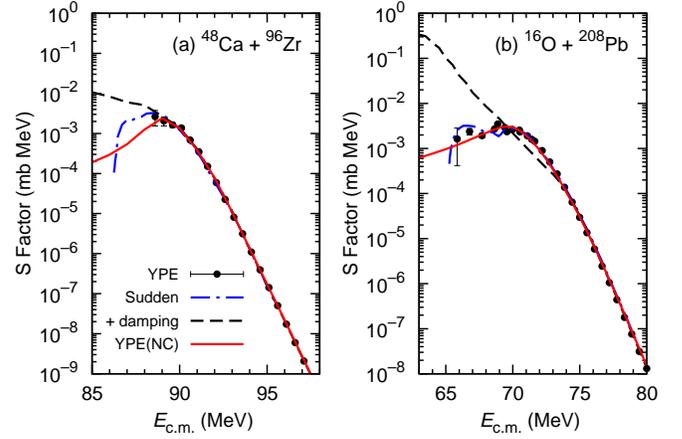}
 \caption{(Color online) Astrophysical S factor representation of the
 fusion cross section versus incident energies for the (a) $^{48}$Ca +
 $^{96}$Zr and (b) $^{16}$O + $^{208}$Pb systems. All symbols are the
 same as those in Fig.~\ref{sfact1}} \label{sfact3}
\end{figure}

Figure \ref{sfact3} shows the astrophysical S factor representations of
the fusion cross sections for the $^{48}$Ca + $^{96}$Zr and $^{16}$O +
$^{208}$Pb systems.  I take $\eta_0=77.0$ and 49.0 for the $^{48}$Ca +
$^{96}$Zr and $^{16}$O + $^{208}$Pb systems, respectively.  The results
calculated with the damping factor (solid lines) are in good agreement
with the experimental data (solid circles). For comparison, the result
calculated with the sudden model taken from Ref.~\cite{Esb10} is plotted
by the dash-dotted line in Fig.~\ref{sfact3} (b).  In the sudden model,
the S factor is suddenly cut off around 66 MeV corresponding to the
bottom of the shallow potential pocket. Alternatively, at low incident
energies, the S factor calculated with the adiabatic model linearly
decreases with decreasing incident energy. The result calculated with
the adiabatic model has a much smoother and weaker energy dependence
than that with the sudden model.

\begin{figure}[htbp]
 \includegraphics[keepaspectratio,width=\linewidth]{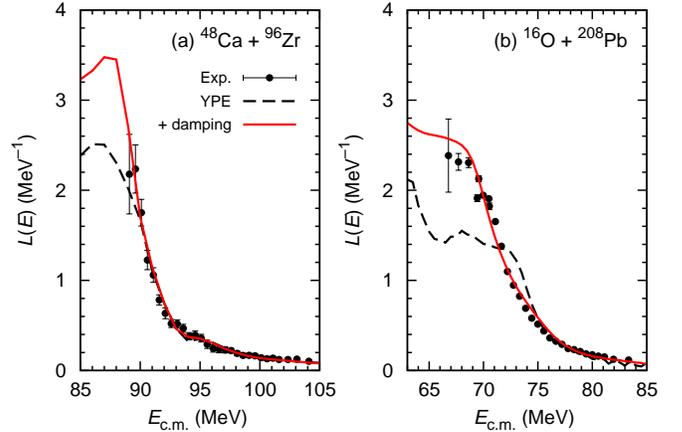}
 \caption{(Color online) Logarithmic derivative representations of the
 fusion cross sections versus incident energies for the (a) $^{48}$Ca +
 $^{96}$Zr and (b) $^{16}$O + $^{208}$Pb systems. All symbols are the same
 as those in Fig.~\ref{log1}.}  \label{log3}
\end{figure}

Figure \ref{log3} shows the logarithmic derivative representations of
the fusion cross sections versus incident energies for the $^{48}$Ca +
$^{96}$Zr and $^{16}$O + $^{208}$Pb systems. The results calculated with
the damping factor (solid lines) are in good agreement with the
experimental data (solid circles). In these systems, the results
calculated with the adiabatic model are saturated at extremely low
incident energies.  Conversely, the results calculated with the sudden
model may significantly increase with decreasing incident energy, as
shown in Ref.~\cite{mis06}.

I also calculated the fusion cross section for the $^{12}$C + $^{198}$Pt
system where the fusion hindrance was recently observed~\cite{Arad15}.
The result calculated with the damping factor well reproduces the
experimental data of the fusion cross section and its S factor and
logarithmic derivative representations. The potential energy at the
touching point also strongly correlates with the incident threshold
energy for fusion hindrance.

\section{Discussion}

\subsection{Adiabatic potential}

\begin{figure}[htbp]
 \includegraphics[keepaspectratio,width=7cm]{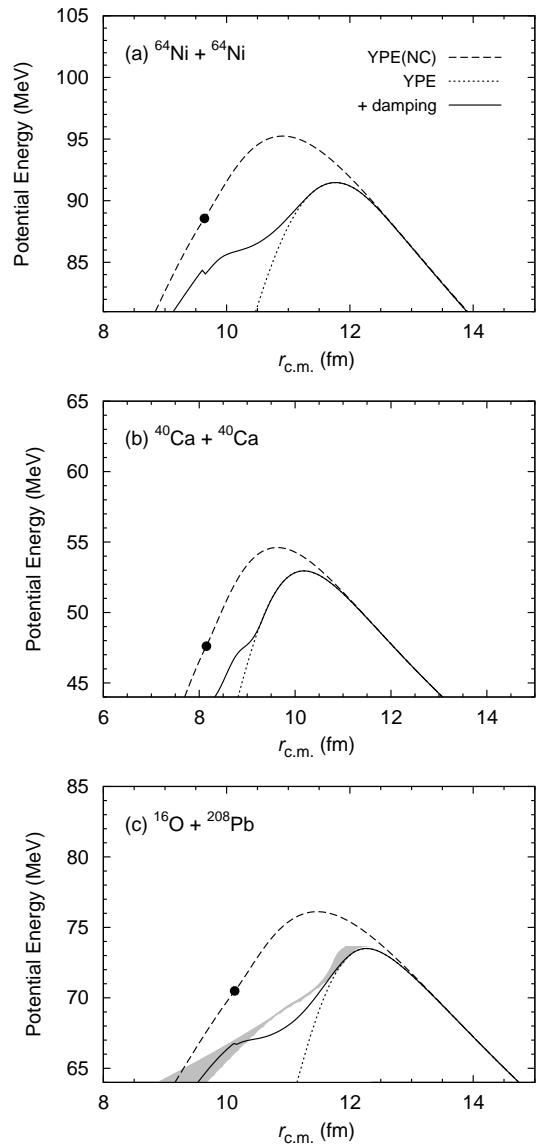}
 \caption{Adiabatic potentials in CC calculations versus the
 center-of-mass distance of colliding nuclei for the (a) $^{64}$Ni +
 $^{64}$Ni, (b) $^{40}$Ca + $^{40}$Ca, and (c) $^{16}$O + $^{208}$Pb
 systems. The solid and dotted lines indicate the results calculated
 with and without the damping factor, respectively. The dashed lines
 indicate the results calculated with no couplings. The solid circles
 denote the energies at the touching point. The gray area in (c)
 represents the adiabatic potential extracted from the potential
 inversion method~\cite{Hagino07}.}  \label{adia}
 \end{figure}

I discuss here the adiabatic potential, namely, the lowest eigenvalue
obtained by diagonalizing the coupling matrix elements in the CC model
at each center-of-mass distance \cite{Hagino07}.  Figure \ref{adia}
shows the adiabatic potential obtained for the (a) $^{64}$Ni +
$^{64}$Ni, (b) $^{40}$Ca + $^{40}$Ca, and (c) $^{16}$O + $^{208}$Pb
systems. The solid and dotted lines indicate the adiabatic potentials
calculated with and without the damping factor, respectively.

In each of these figures, one can see that the adiabatic potential
calculated with the damping factor (solid line) becomes thicker than
that without the damping factor (dotted line) below the potential energy
at the touching point indicated by the solid circle. This is the main
effect of the damping factor on the fusion hindrance behavior.  In the
adiabatic model, the adiabatic potential becomes thicker below the
potential energy at the touching point and this increase in thickness
naturally accounts for the fusion hindrance behavior.

In Fig.~\ref{adia} (b), it seems that the increase in the thickness of
the adiabatic potential is relatively small compared to the other
heavy-mass compound systems.  In fact, the fusion hindrance behavior of
the fusion cross sections obtained with the adiabatic model is small in
the medium-light mass system.  This is associated with coupling
strengths causing the enhancements in the fusion cross sections.  In
this system, the difference between the thicknesses of the bare
potential (dashed line) and the adiabatic potential without the
damping factor (dotted line) is small, indicating that the coupling
strengths in this system are weak compared to those of heavier compound
mass systems.  Thus, the increase in the thickness of the adiabatic
potential due to the damping factor also becomes small. As a result, the
fusion hindrance behavior is relatively small in the medium-light mass
systems.

In Fig.~\ref{adia} (c), the adiabatic potential extracted with the
potential inversion method \cite{Hagino07} from the experimental data is
illustrated by the gray area. Clearly, the adiabatic potential
calculated with the damping factor strongly correlates with that of the
potential inversion method. This is strong evidence for the presence of
the smooth transition from sudden to adiabatic processes and the
coordinate-dependent coupling strength.

\subsection{Barrier Distribution}

\begin{figure}[htbp]
 \includegraphics[keepaspectratio,width=7cm]{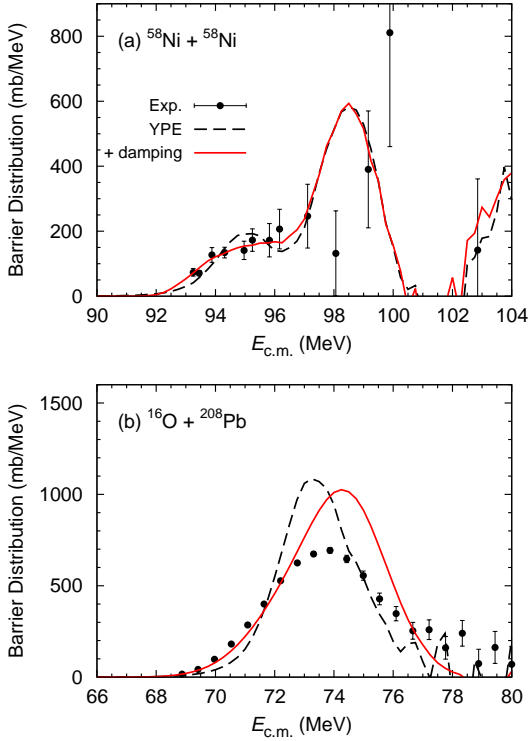} \caption{(Color
 online) Barrier distribution versus incident energies for the (a)
 $^{58}$Ni + $^{58}$Ni and (b) $^{16}$O + $^{208}$Pb systems. The solid
 and open circles denote the experimental data. The solid and dashed
 lines indicate the calculated results with and without the damping
 factor, respectively. } \label{bdis}
 \end{figure}

For the $^{58}$Ni + $^{58}$Ni and $^{16}$O + $^{208}$Pb systems, a large
improvement in the barrier distribution $D_{\rm fus}(E)=d^2(E\sigma_{\rm
fus})/dE^2$ has been achieved by taking into account the damping factor.
Figure \ref{bdis} shows the barrier distributions versus incident
energies for the (a) $^{58}$Ni + $^{58}$Ni and (b) $^{16}$O +
$^{208}$Pb systems.

In Fig.~\ref{bdis} (a), one can see a small peak of the barrier
distribution calculated without the damping factor (dashed line) around
$E_{\rm c.m.}=95$ MeV.  By taking into account the damping factor, this
peak structure vanishes, and the plateau structure appears between about
94-97 MeV (solid line). This plateau structure is in good agreement with
the experimental data (solid circles).  The experimental data for this
system is adequate for investigating the properties of the fusion
hindrance, because a clear signature of the fusion hindrance appears in
the barrier distribution of the fusion cross sections around
$10^{-1}\sim1$ mb.  In addition, the strong fusion hindrance can be seen
for this system in the calculated result of the fusion cross section
after implementing the damping factor.  However, the accuracy of the
experimental data is still insufficient to determine the energy
dependence of the calculated barrier distribution.  Thus, in this
system, much higher precision fusion data around these incident energies
are required to study the fusion hindrance properties in details.

\begin{figure}[htbp]
 \includegraphics[keepaspectratio,width=7cm]{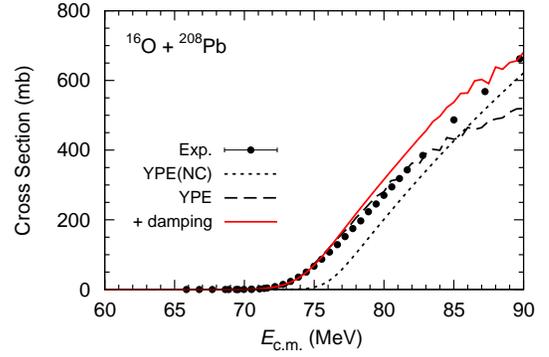} \caption{(Color
 online) Fusion cross section versus incident energies for the $^{16}$O
 + $^{208}$Pb system. The solid circles denote the experimental
 data. The solid and dashed lines indicate the results calculated with
 and without the damping factor, respectively. The dotted line indicates
 the calculated result with no couplings.}  \label{crosshigh}
 \end{figure}

For the $^{16}$O + $^{208}$Pb system, the agreement between the tails of
the experimental data and the result calculated with the damping factor
in the low incident-energy region is greatly improved [solid and dashed
lines in Fig.~\ref{bdis} (b)].  The peak position of the result
calculated with the damping factor shifts to the higher incident energy
by about 2 MeV, compared to that without the damping factor.  The
calculated barrier distribution at the peak position is also reduced by
taking into account the damping factor.  However, it seems that the
width of the barrier distribution calculated with the damping factor
becomes wider than that without the damping factor.  In fact, for this
system, fusion cross sections at incident energies highly above the
Coulomb barrier are overestimated when the damping factor is employed
(solid line in Fig.~\ref{crosshigh}).  In this system, other dissipative
mechanisms, including single particle excitations, as discussed in
Refs.~\cite{Yusa10,Yusa13}, are necessary to reproduce the fusion cross
sections at incident energies highly above the Coulomb barrier.  The YPE
potential, which is optimized in the adiabatic process, may be also
inapplicable to this high incident-energy region.  Note that in the
systems except for the mass-asymmetric $^{16}$O + $^{208}$Pb and
$^{12}$C + $^{198}$Pt systems, the calculated fusion cross sections with
the adiabatic model are in good agreement with the experimental data
even above the Coulomb barrier.

\subsection{Average angular momentum of compound nucleus}

\begin{figure}[htbp]
 \includegraphics[keepaspectratio,width=7cm]{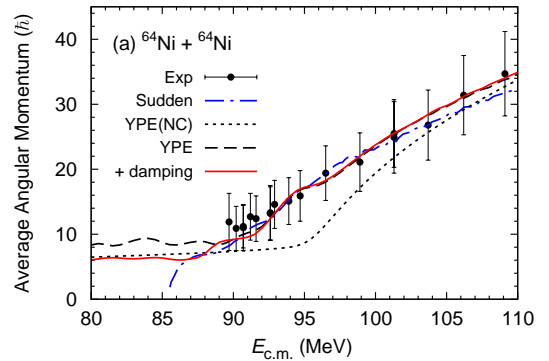} \caption{(Color
 online) Average angular momentum of compound nucleus versus incident
 energies for the $^{64}$Ni + $^{64}$Ni system. The sold circles denote
 the experimental data from Ref.~\cite{Ack96}. The solid and dashed
 lines indicate the calculated results with and without the damping
 factor, respectively. The dotted lines indicate the CC calculation with
 no couplings. The dash-dotted line indicates the result calculated with
 the sudden model taken from Ref.~\cite{mis06}.}  \label{avang}
 \end{figure}
\begin{figure}[htbp]
 \includegraphics[keepaspectratio,width=7cm]{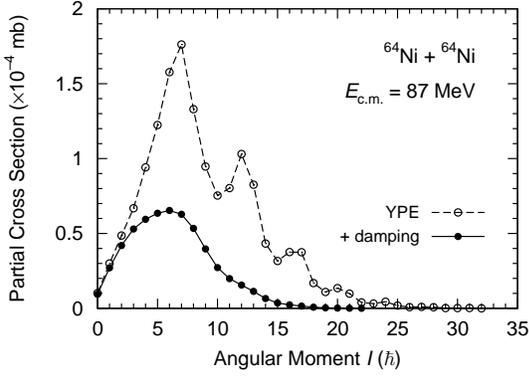} \caption{Partial
fusion cross section versus angular momentum $I$ at an incident energy
of 87 MeV for the $^{64}$Ni + $^{64}$Ni system. The solid line with
solid circles and the dashed line with open circles indicate the results
calculated with and without the damping factor, respectively.}
\label{pcross}
 \end{figure}

An important piece of experimental data for discriminating between the
adiabatic and sudden models is the average angular momentum of a
compound nucleus at extremely low incident energies. Figure \ref{adia}
shows the average angular momentum of the compound nucleus versus
incident energies for the $^{64}$Ni + $^{64}$Ni system.  The experimental data for the $^{64}$Ni + $^{64}$Ni
is from Ref.~\cite{Ack96}, denoted by the solid circles.
The solid and dashed lines indicate the results calculated with and
without the damping factor, respectively. In the $^{64}$Ni + $^{64}$Ni
system, the result calculated with the damping factor decreases with
decreasing incident energy. Below the potential energy at the touching
point ($V_{\rm Touch}=88.6$ MeV), the result calculated with the damping
factor remains constant at around 8 $\hbar$. Subsequently, the result
calculated with the damping factor is lower than that without the
damping factor by about 20\%.
For the $^{12}$C + $^{198}$Pt system, the
result calculated with the damping factor is in good agreement with the
experimental data of overall incident energy~\cite{Arad15}.

Figure \ref{pcross} shows the calculated partial cross sections versus
the angular momentum $I$ at an incident energy of 85 MeV for the
$^{64}$Ni + $^{64}$Ni system.  The solid line with solid circles and the
dashed line with open circles indicate the results calculated with and
without the damping factor, respectively.  In the figure, in the
adiabatic model, the partial cross section at each $I$ is naturally
reduced by the effects of the damping factor.  On the other hand, the
average angular momentum calculated with the sudden model is strongly
suppressed at energies below the threshold energy [dash-dotted line in
Fig.~\ref{avang} (a)].  This result indicates that a mechanism of the
fusion hindrance in the sudden model would be the cutoff of high
angular-momentum components in the partial cross sections due to the
shallow potential pocket.  In this respect, a compound nucleus formed at
the deep sub-barrier incident energy in the sudden model has
substantially low angular momentum.  Subsequently, particle evaporation
from the compound state would be forbidden, because the angular momentum
necessary for particle evaporation is insufficient.  Therefore, the
properties of the formed compound nucleus are considerably different
between the adiabatic and sudden models.  Thus, in order to discriminate
the two models, it is also important to measure the average angular
momentum at deep sub-barrier energies.
 
\subsection{Systematic trends of the radius and diffuseness parameters in the damping factor}

\begin{figure}[htbp]
 \includegraphics[keepaspectratio,width=7cm]{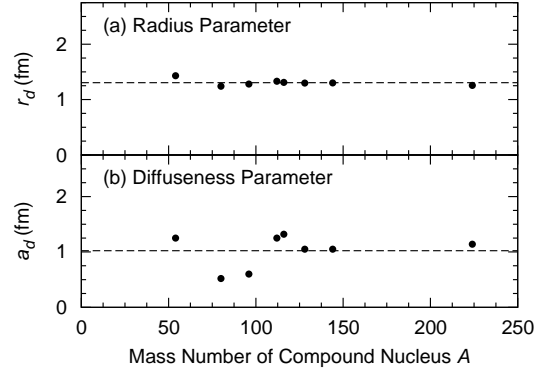}
 \caption{Systematic trends of the (a) radius and (b) diffuseness
 parameters in the damping factor versus the mass numbers of the
 compound nuclei for the systems presented in this paper. The solid
 circles denote the obtained values fitted to the experimental data. The
 dashed lines indicate the average value of all the obtained values in
 individual radius and diffuseness parameters.}  \label{sysfact}
\end{figure}

Next, I test the systematic trends of the obtained radius and
diffuseness parameters in the damping factor for the systems presented
in this paper. Figure \ref{sysfact} shows the (a) radius and (b)
diffuseness parameters versus the mass numbers of the compound nuclei
for the systems presented in this paper. In the figure, the solid
circles denote the obtained values. The dashed line indicates the
average value for all the obtained values of the individual radius and
diffuseness parameters.  Clearly, the values of $r_d$ are almost
constant at around an average value of 1.31 fm.  Except for a few
points, the values of $a_d$ are also distributed around an average value
of 1.02 fm.

As shown in Refs.~\cite{ich13,ich15}, the damping factor strongly
correlates with the damping of the transition strength of individual
colliding nuclei when they approach one another. In this respect, the
damping of the transition strengths would start at the overlapping
between the tails of the density distributions for colliding
nuclei. That is, the radius parameter of the damping factor would
correlate with a range of interactions between the colliding
nuclei. This would result in an almost constant value of $r_d$ in all
the systems. Alternatively, $a_d$ is associated with the damping
strength of quantum vibrations, which would strongly depend on an inner
nuclear structure of individual colliding nuclei. In fact, the values
of $a_d$ for the $^{40}$Ca + $^{40}$Ca and $^{48}$Ca + $^{48}$Ca
systems, where $^{40,48}$Ca have the strong shell effects, which largely
deviate from the average value. In this sense, the values of $a_d$ would
be somewhat scattered around the average value.

\subsection{Correlation between $V_{\rm Touch}$ and the incident threshold energy for fusion hindrance}

As shown in Sec.~III, the potential energies at the touching point
$V_{\rm Touch}$ strongly correlate with the incident threshold energies
for fusion hindrance in the medium-heavy mass and mass-asymmetric
reaction systems (see the arrows in Figs.~\ref{cross1} and
\ref{cross3}).  In the medium-light mass systems, the correlation is
relatively weak. This would result from considerably larger deformation
parameters associated with coupling strengths in these systems compared
to those in the other systems.  In addition, the curvature of the bare
potentials around the Coulomb barrier is relatively large
[Fig.~\ref{adia} (b)]. Thus, the effect of the damping factor starts
much before the touching point in this system.  In this regards, the
threshold rule should be modified considering the deformation parameter
effects in this system.  This modification is now in progress.

\section{Summary}

In summary, I have proposed a novel extension of the standard CC model
to describe the fusion hindrance phenomenon observed at extremely low
incident energies.  I have systematically investigated 
various deep sub-barrier fusion reactions using an adiabatic approach.

A key quantity for understanding fusion hindrance is the potential
energy at the touching point $V_{\rm Touch}$. The threshold incident
energies for fusion hindrance strongly correlate with $V_{\rm
Touch}$. At incident energies below $V_{\rm Touch}$, the inner turning
point of the potential energy becomes inside the touching point. That
is, a composite system must penetrate through a residual Coulomb
potential with an overlapping configuration before fusion occurs.  Thus,
the dynamics in the overlapping region of the two colliding nuclei would
be responsible for the fusion hindrance.

I have described the fusion hindrance based on an adiabatic approach.
In the adiabatic approach, fusion is assumed to occur slowly, and neck
formation occurs in the overlapping region. The nuclear density
distributions then evolve with the lowest-energy configuration. Based on
this picture, one can calculate the one-body potential energy with an
appropriate adiabatic model. However, one cannot directly apply the
obtained one-body potential to a standard CC model, because of double
counting of CC effects.

To avoid this double counting, an important extension of the standard CC
model is the introduction of a damping factor in the coupling form
factor, which enables a smooth transition from sudden to 
adiabatic processes. Namely, the damping factor represents the damping
of quantum vibrations and suppresses transitions between reaction
channels when two colliding nuclei approach one another.  By introducing
the damping factor in the coupling form factor, one can correctly
estimate the tunneling probability in the one-body region, when an
appropriate one-body potential is taken into account in the bare heavy
ion-ion potential in the CC model.

In this paper, I adopted the YPE model as a basic heavy ion-ion
potential.  An advantage of this potential model is a unified
description of both the two- and one-body systems.  For the purpose of
systematic investigations, in this paper, the one-body potential is
approximated with a third-order polynomial function.  This procedure
works very well, because the lowest incident energies in the
experiments, which have been performed until now, are lower than the
potential energies at the touching point only by a few MeV.

Based on this framework, I have systematically calculated the
medium-heavy mass systems of $^{64}$Ni + $^{64}$Ni, $^{58}$Ni +
$^{58}$Ni, and $^{58}$Ni + $^{54}$Fe, the medium-light mass systems of
$^{40}$Ca + $^{40}$Ca, $^{48}$Ca + $^{48}$Ca, and $^{24}$Mg + $^{30}$Si,
and the mass-asymmetric systems of $^{48}$Ca + $^{96}$Zr, $^{12}$C +
$^{198}$Pt, and $^{16}$O + $^{208}$Pb. In addition, I have calculated
their fusion cross sections, the astrophysical S factor and the
logarithmic derivative representations of those.  In all the systems,
the calculated results are in excellent agreement with the experimental
data, indicating that the smooth transition from sudden to adiabatic
processes occurs in the deep sub-barrier fusion reactions, and the
coordinate-dependent coupling strength is responsible for the fusion
hindrance.

I have also showed the adiabatic potential, which is the lowest
eigenvalue obtained by diagonalizing the coupling matrix elements, in the
CC model for the $^{64}$Ni + $^{64}$Ni, $^{40}$Ca + $^{40}$Ca, and
$^{16}$O + $^{208}$Pb systems. Because of the introduction of the damping
factor, the adiabatic potential using the damping factor becomes much
thicker than without the damping factor. This naturally accounts
for the fusion hindrance behavior. The adiabatic potential for the
$^{16}$O + $^{208}$Pb systems is in good agreement with that extracted
from the potential inversion method.  For the $^{58}$Ni + $^{58}$Ni and
$^{16}$O + $^{208}$Pb systems, large improvements in the barrier
distribution at low incident energies were made by
taking into account the damping factor.

The adiabatic and sudden models significantly differ in the calculated
average angular momentum of the compound nuclei.  The average angular
momentum estimated with the adiabatic model is saturated below the
threshold incident energy. Conversely, that with the sudden model
is strongly suppressed at low incident energies, because a mechanism of
the fusion hindrance in the sudden model is the cutoff of high
angular-momentum components in the partial waves due to the shallow
potential pocket.

The obtained parameters of individual $r_d$ and $a_d$ in the damping
factor are nearly constant in all the systems. The values of $a_d$ are
somewhat scattered around its average value, because $a_d$ would depend
on the shell structure of the composite system.  The strong correlation
between $V_{\rm Touch}$ and the threshold incident energy for the fusion
hindrance can be seen in the medium-heavy mass and mass-asymmetric
systems.  For the medium-light mass system, this correlation is somewhat
weak, because the deformation parameters associated with the coupling
strengths in this system are large compared to those at heavier compound
mass systems.  It is necessary to modify the threshold rule with the
effects of the deformation parameters in the medium-light mass systems.

\begin{acknowledgments}
 The author thanks K. Hagino for his helpful advice and useful
 discussions. The author also thanks A. Iwamoto and K. Matsuyanagi for
 useful discussions.  This work was partially supported by the MEXT HPCI
 STRATEGIC PROGRAM and by JPSJ KAKENHI Grant Number 15K05078.
\end{acknowledgments}

\appendix
\section{Lemniscatoids parametrization}

\begin{figure}[htbp]
 \includegraphics[keepaspectratio,width=6cm]{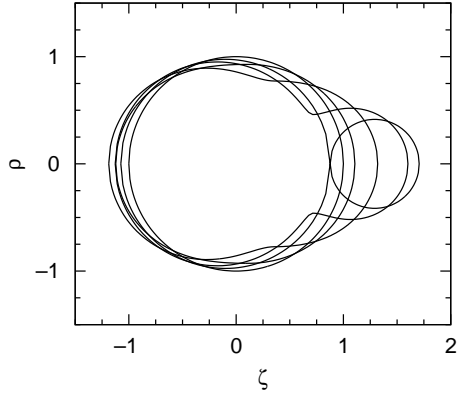} \caption{Shapes
 described using the lemniscatoids parametrization from $s=0$ to 1
 with an increment of 0.25. The mass asymmetry $\alpha$ of the shape
 corresponds to the $^{16}$O + $^{208}$Pb system.}  \label{lemni}
\end{figure}

The lemniscatoids parametrization proposed by Royer
\cite{Roy82,Roy84,Roy85} is a special case of the Cassinian oval.  This
parametrization allows us to smoothly describe from the touching
configuration of two spheres to a single spherical shape as functions of
the elongation parameter $s$ and the mass-asymmetry parameter $\alpha$.
The described shape has a single sphere at $s=0$ and the touching
configuration of two spheres at $s=1$.  The mass-asymmetry
parameter is defined as $\alpha=(A_1 - A_2)/(A_1 + A_2)$, where $A_1$
and $A_2$ are the total mass numbers of the fragments 1 and 2. The radius
of the spherical composite system is given by $R_0=r_0(A_1 +
A_2)^{1/3}$, where $r_0$ is the radius parameter.

In the cylindrical coordinate system, the radial $\rho$ and $z$
coordinates are expressed as the dimensionless parameters
$\varrho=\rho/R_0$ and $\zeta=z/R_0$. The radial displacement of the
shapes described with the lemniscatoids parametrization for the
fragments 1 and 2 is given by
\begin{equation}
 \varrho_i^2(\zeta)=
  \begin{cases}
 \displaystyle
   \frac{1}{2}\left(A^2-2\zeta_c^2 + \sqrt{A^4+4(C_i^2-A^2)\zeta_c^2}\right)& (A\ne 0) \\[16pt]
   C_i|\zeta_c|-\zeta_c^2& (A=0),
  \end{cases}
\end{equation}
where $\zeta_c=\zeta - \zeta_g$, $\zeta_g$ is the constant to shift the
center-of-mass position of the whole system to the origin, $A$ is the
neck diameter parameter, and $C_i$ are the radius parameter of
individual fragments 1 and 2. The index $i$ stands for fragments 1 and
2.  The parameters $A$ and $C_i$ are determined by the condition of the
volume conservation. The regions of $z$ for the fragments 1 and 2 are
given by $-C_1 + \zeta_g \leq \zeta \leq \zeta_g $ and $ \zeta_g \leq
\zeta \leq C_2 + \zeta_g$, respectively.

Here, the parameter $S_i$ are defined as $S_i=A/C_i$. For $\alpha>0$,
Royer assumed the $s$ dependence of $S_i$ as $S_1=s$ and $S_2=S_1/f_s$
\cite{Roy84,Roy85}, where
\begin{equation}
 f_s=\sqrt{s^2+(1-s^2)\left(\frac{1-\alpha}{1+\alpha}\right)^{2/3}}.
\end{equation}
From this definition, one also obtains $C_2=f_sC_1$. The volume of each
fragment $V_i$ is given by $V_i/R_0^3=C_i^3v_i$, where
\begin{align}
 v_i&=\frac{\pi}{2}\left(S_i^2-\frac{2}{3}+\int_0^1\sqrt{4(1-S_i^2)\zeta^2+S_i^4}d\zeta\right) \\
 &=\frac{\pi}{24}\left(4+6S_i^2+\frac{3S_i^4}{\sqrt{1-S_i^2}}\text{Arcsinh}\left(2\frac{\sqrt{1-S_i^2}}{S_i^2}\right)\right).
\end{align}
For practical calculations, Eq.~(A3) and the numerical integration for
its last term are conveniently used to avoid the divergence of Eq.~(A4)
at $S_i=0$ and $1$.  The total volume $V$ is given by
$V/R_0^3=V_1/R_0^3+V_2/R_0^3=C_1^3v_1+C_2^3v_2$. With $C_2=f_sC_1$, one
obtains $V/R_0^3=C_1^3(v_1+f_s^3v_2)$. Since $V=\frac{4}{3} \pi R_0^3$,
$C_1$ is given by
\begin{equation}
 C_1^3=\frac{4}{3}\pi\frac{1}{v_1+f_s^3v_2}.
\end{equation}
Then, $A$ and $C_2$ are calculated by $A=S_1C_1$ and $C_2=f_sC_1$. For
$\alpha<0$, take $S_1=S_2/f_s$ and $S_2=s$ with $|\alpha|$ and exchange
the indexes 1 and 2 in the above equations.

Next, I calculate $\zeta_g$ and the center-of-mass distance
 $r$ between the fragments 1 and 2.  With $\zeta_g=0$ in Eq.~(A1),
 $\zeta_g$ is given by
\begin{equation}
 \zeta_g = -\frac{3}{4}\left(\int_{-C_1}^0\zeta\varrho_1^2(\zeta)d\zeta+\int_0^{C_2}\zeta\varrho_2^2(\zeta)d\zeta\right). 
\end{equation}
Again with $\zeta_g=0$ in Eq.~(A1), $r$ is given by
\begin{equation}
 \frac{r}{R_0} = \frac{\pi}{C_2^3v_2}\int_0^{C_2}\zeta\varrho_2^2(\zeta)d\zeta-\frac{\pi}{C_1^3v_1}\int_{-C_1}^0\zeta\varrho_1^2(\zeta)d\zeta. 
\end{equation}

Figure \ref{lemni} shows the shapes described using the
lemniscatoids parametrization from $s=0$ to 1 with an increment of 0.25.
The mass asymmetry corresponds to the $^{16}$O + $^{208}$Pb system.
From the figure, a smooth transition can be seen from the single
spherical shape to where the two spherical shapes touch.

The first and second derivatives of Eq.~(A1) are necessary for
calculating Eq.~(\ref{ypeint}).  The first derivative is given by
\begin{equation}
\frac{d\varrho_i^2(\zeta)}{d\zeta}=
  \begin{cases}
   \displaystyle
   -2\zeta_c\left(\frac{2(\varrho_i^2+\zeta_c^2)-C_i^2}{2(\varrho_i^2+\zeta_c^2)-A^2}\right)&  (A\ne 0)\\[16pt]
   \text{sign}(\zeta_c)C_i-2\zeta_c& (A=0).
  \end{cases}
\end{equation}
The second derivative is given by
\begin{widetext}
\begin{equation}
\frac{d^2\varrho_i^2(\zeta)}{d\zeta^2}=
  \begin{cases}
   \displaystyle
   \frac{-2}{2(\varrho_i^2+\zeta_c^2)-A^2}\left(2(\varrho_i^2+\zeta_c^2)-C_i^2+\left(\frac{d\varrho_i^2}{d\zeta}+2\zeta_c\right)^2\right)&  (A\ne 0)\\[16pt]
   -2& (A=0).
  \end{cases}
\end{equation}
\end{widetext}

\section{Approximation of the one-body potential energy}

\begin{figure}[b]
 \includegraphics[keepaspectratio,width=7cm]{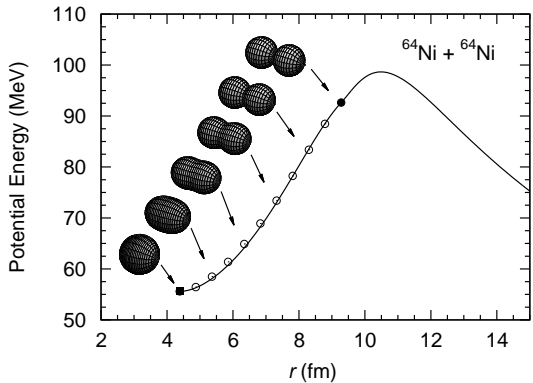}
 \caption{Comparison between the approximation with a third-order
 polynomial function and the liquid drop energies calculated with the
 lemniscatoids parametrization. The solid curve indicates the
 calculated result of the approximation. The open circles denote the
 liquid drop energies with the lemniscatoids parametrization. Some of
 nuclear shapes described with the lemniscatoids parametrization are
 displayed.}  \label{comp}
\end{figure}

Here I describe how to construct the adiabatic one-body potential used in
this paper.  For simplicity, I approximate the one-body potential energy
with a third-order polynomial function based on the YPE model for
the purpose of systematic investigations.
Thus, the one-body nuclear potential energy $V_{\rm
1bd}^{(N)}$ is given by
\begin{equation}
V_{\rm 1bd}^{(N)}(r) = C_0+C_1r+C_2r^2 +C_3r^3,
\end{equation}
where $C_n$ are coefficients determined by the condition that $V_{\rm
1bd}^{(N)}$ is smoothly jointed to the two-body nuclear potential,
$V_{\rm 2bd}^{(N)}$, at the touching point, $r_{\rm T}$. I impose
that the values of $V_{\rm 1bd}$ and $V_{\rm 2bd}$ and the first
derivatives of those are smoothly jointed at the touching point.  To
determine $C_n$, I also assume the position of the compound state
$r_{\rm GS}$, the energy of the compound state $E_{\rm GS}$, and the
curvature of the potential energy at the compound state
$\hbar\omega_{\rm GS}$. The curvature $\hbar\omega_{\rm GS}$ corresponds
to the parabolic approximation of the potential energy at the compound
state given by $V_{\rm 1bd}\simeq\frac{1}{2}\mu\omega_{\rm GS}^2r^2$
$(r\sim r_{\rm GS})$, where $\mu$ is the reduced mass of the reaction
system. For the Coulomb potential, I use the point charge approximation
given by $V_{\rm 1bd}^{(C)}(r)=Z_TZ_Pe^2/r$.  These conditions are
expressed in the following matrix form:
\begin{equation}
\begin{pmatrix}
 1 & r_{\rm T} & r_{\rm T}^2 &r_{\rm T}^3 \\
 0 & 1 & 2r_{\rm T}&3r_{\rm T}^2\\
 1 & r_{\rm GS} & r_{\rm GS}^2&r_{\rm GS}^3\\
 0 & 0 & 2 & 6r_{\rm GS}
\end{pmatrix}
 \begin{pmatrix}
  C_0 \\ C_1 \\ C_2 \\ C_3
 \end{pmatrix}
 =
 \begin{pmatrix}
  V_{\rm 2bd}^{(N)}(r_{\rm T}) \\
  {V_{\rm 2bd}^{(N)}}'(r_{\rm T}) \\
  E_{\rm GS} - V_{\rm 1bd}^{(c)}(r_{\rm GS}) \\
  \mu\omega_{\rm GS}^2-{V_{\rm 1bd}^{(c)}}''(r_{\rm GS})
 \end{pmatrix}.
\end{equation}
I numerically solve these linear equations and obtain the values of $C_n$.

To verify the performance of this procedure, I calculate the liquid-drop
energy with Eq.~(13) using the lemniscatoids parametrization which
describes appropriate neck formations \cite{Roy82,Roy84} in the
$^{64}$Ni + $^{64}$Ni system and compare the approximation of the
third-order polynomial function with those.  In this parametrization,
nuclear shapes are described as a function of the center-of-mass
distance between two halves of the composite system (Appendix A).
To obtain the potential energy, I subtract the self-volume energies of
the two colliding nuclei from the total liquid drop energy. In this
calculation, I use $r_0=1.16$ fm and $a=0.68$ fm. The Coulomb volume
energy is also calculated with shape configurations given by the
lemniscatoids parametrization.  In the procedure described above, the
three input parameters $E_{\rm GS}$, $r_{\rm GS}$, and
$\hbar\omega_{\rm GS}$ are estimated from the liquid drop energy
calculated with the lemniscatoids parametrization. These are taken as
$E_{\rm GS}=55.60$ MeV, $r_{\rm GS}=4.38$ fm, and $\hbar\omega_{\rm
GS}=3.00$ MeV.

\begin{figure}[b]
 \includegraphics[keepaspectratio,width=7cm]{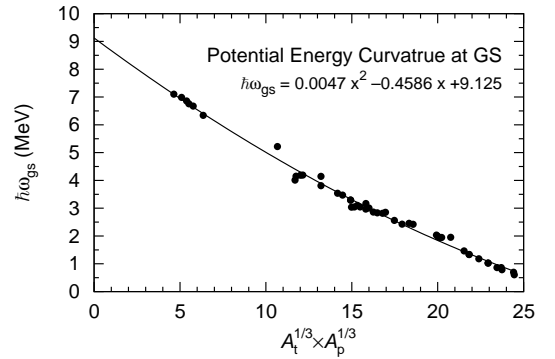}
 \caption{The values of $\hbar\omega_{\rm GS}$ calculated with the
 liquid-drop model using the lemniscatoids parametrization versus
 $x=A_T^{1/3}\cdot A_P^{1/3}$ for various systems. The solid circles
 denote the calculated results. The solid curve represents a curve
 fitted to the calculated results.}  \label{homg}
\end{figure}

Figure \ref{comp} shows the calculated result of the approximation with
the third-order polynomial function (solid lines) and the the liquid
drop energy calculated with the lemniscatoids parametrization (open
circles). In the figure, the solid circle and square denote the
potential energy at the touching point and the energy of the compound
state calculated with the liquid drop model, respectively. Some nuclear
shapes described with the lemniscatoids parametrization are displayed in
the figure.  In the figure, the solid curve is clearly in good agreement
with the open circles, indicating that the approximation performs very
well.

In the procedure described above, there are the three input parameters
$E_{\rm GS}$, $r_{\rm GS}$, and $\hbar\omega_{\rm GS}$.  In this paper,
I estimate the value of $E_{\rm GS}$ from the energy of the compound
state calculated with experimental nuclear masses, taken from the
AME2003 table \cite{audi03}.  The value of $r_{\rm GS}$ is calculated
with the lemniscatoids parametrization. In this parametrization, the
compound state always exhibits a spherical configuration. Thus, $r_{\rm
GS}$ is defined as the center-of-mass distance between two halves of the
spherical nucleus. This is given by $r_{\rm GS}=3/4\cdot R_c$, where
$R_c$ denotes the nuclear radius of the compound state with
$R_c=r_0(A_T+A_P)^{1/3}$.

The value of $\hbar\omega_{\rm GS}$ is estimated from a fitting curve
obtained by systematically investigating the liquid drop energy using
the lemniscatoids parametrization (Appendix A).  I calculate $V_{\rm
1bd}''$ at the compound state using the liquid drop model with the
lemniscatoids parametrization for the reaction systems shown in
Ref.~\cite{Jiang06}. I also calculate those for some cold fusion
reactions with the $^{208}$Pb target.  Figure \ref{homg} shows the
calculated values of $\hbar\omega_{\rm GS}$ as a function of
$x=A_T^{1/3}\cdot A_P^{1/3}$. In the figure, a strong correlation
between $\hbar\omega_{\rm GS}$ and $x$ is seen. I fitted the obtained
values with a second-order polynomial function. The fitted curve is
given by $\hbar\omega_{\rm GS}=0.0047x^2-0.4586x+9.125$ MeV, which is
presented by the solid curve in the figure.


\begin{thebibliography}{50}
  \bibitem{DHRS98} M. Dasgupta, D.J. Hinde, N. Rowley, and A.M. Stefanini,
	  Annu. Rev. Nucl. Part. Sci. {\bf 48}, 401 (1998).
  \bibitem{Bal98}
	  A. B. Balantekin and N. Takigawa, Rev. Mod. Phys. {\bf 70}, 77
	  (1998).
  \bibitem{Hagino12}
	  K. Hagino and N. Takigawa, Prog. Theor. Phys. {\bf
	  128}, 1061 (2012).
  \bibitem{Jiang02} C. L. Jiang, H. Esbensen, K. E. Rehm, B. B. Back,
	  R. V. F. Janssens, J. A. Caggiano, P. Collon, J. Greene,
	  A. M. Heinz, D. J. Henderson, I. Nishinaka, T. O. Pennington,
	  and D. Seweryniak, Phys. Rev. Lett. {\bf 89}, 052701 (2002).
  \bibitem{Jiang04}
	  C. L. Jiang {\it et al.}, Phys. Rev. Lett. {\bf 93}, 012701
	  (2004).
  \bibitem{Jiang06} C. L. Jiang, B. B. Back, H. Esbensen, R. V. F. Janssens, and K. E. Rehm, Phys. Rev. C 73, 014613 (2006).
  \bibitem{Back14}
	  B.B. Back {\it et al.}, Rev. Mod. Phys. {\bf 86}, 317 (2014).

  \bibitem{ich07-2}
	  T. Ichikawa, K. Hagino, and A. Iwamoto, Phys. Rev. C {\bf 75},
	  064612 (2007).

  \bibitem{mis06}
	  \c{S} Mi\c{s}icu and H. Esbensen, Phys. Rev. Lett. {\bf 96},
	  112701 (2006); Phys. Rev. C
	  {\bf 75}, 034606 (2007).

  \bibitem{Esb07}
	  H. Esbensen and \c{S} Mi\c{s}icu, Phys. Rev. C {\bf 76},
	  054609 (2007).
  \bibitem{Monta12} G. Montagnoli, A.M. Stefanini, C.L. Jiang,
	  H. Esbensen, L. Corradi, S. Courtin, E. Fioretto, A. Goasduff,
	  F. Haas, A.F. Kifle, C. Michelagnoli, D. Montanari,
	  T. Mijatovi\ifmmode \acute{c}\else \'{c}\fi{}, K. E. Rehm,
	  R. Silvestri, Pushpendra P. Singh, and F. Scarlassara,
	  S. Szilner, X.D. Tang, and C.A. Ur, Phys. Rev. C {\bf 85},
	  024607 (2012).
  \bibitem{Esb10}
	  H. Esbensen, C. L. Jiang, and A. M. Stefanini,
	  Phys. Rev. C {\bf 82}, 054621 (2010).
  \bibitem{mis11}
	  \c{S}erban Mi\c{s}icu and Florin Carstoiu, Phys. Rev. C {\bf 83}, 054622 (2011).
  \bibitem{Jiang14}
	  C. L. Jiang, A. M. Stefanini, H. Esbensen, K. E. Rehm,
	  S. Almaraz-Calderon, B. B. Back, L. Corradi, E. Fioretto,
	  G. Montagnoli, F. Scarlassara, D. Montanari, S. Courtin,
	  D. Bourgin, F. Haas, A. Goasduff, S. Szilner, and
	  T. Mijatovic, Phys. Rev. Lett. {\bf 113}, 022701 (2014).
  \bibitem{ich07-1}
	  T. Ichikawa, K. Hagino, and A. Iwamoto, Phys. Rev. C {\bf 75},
	  057603 (2007).
  \bibitem{ich09}
	  T. Ichikawa, K. Hagino, A. Iwamoto, Phys. Rev. Lett. {\bf 103},
	  202701 (2009); EPJ Web Conf. {\bf 17}, 07001 (2011).
  \bibitem{ich13}
	  T. Ichikawa, K. Matsuyanagi, Phys. Rev. C {\bf 88},
	  011602(R) (2013).
  \bibitem{ich15}
	  T. Ichikawa, K. Matsuyanagi, Phys. Rev. C {\bf 92}, 021602(R) (2015).
  \bibitem{Umar06} A.S. Umar and V.E. Oberacker, Phys. Rev. C {\bf 74},
          021601 (R) (2006).
  \bibitem{Umar09}
          A.S. Umar and V.E. Oberacker, Eur. Phys. J. A {\bf 39}, 243 (2009).
  \bibitem{Umar14} A.S. Umar, C. Simenel, and V.E. Oberacker,
          Phys. Rev. C {\bf 89}, 034611 (2014).
  \bibitem{Keser12}
          R. Keser, A.S. Umar, and V.E. Oberacker, Phys. Rev. C {\bf 85}, 044606 (2012).
  \bibitem{Esb87} H. Esbensen and S. Landowne, Phys. Rev. C {\bf 35},
	  2090 (1987).

  \bibitem{ccfull}
	  K. Hagino, N. Rowley, and A.T. Kruppa, Comp. Phys. Comm. 123, 143 (1999).
  \bibitem{kra79}
	  H.J. Krappe, J.R. Nix, and A.J. Sierk, Phys. Rev. C {\bf 20}, 992 (1979).

  \bibitem{mo04}
	  P. M\"oller, A.J. Sierk, and A. Iwamoto, Phys. Rev. Lett. 92, 072501 (2004).

  \bibitem{Roy82}
	  G. Royer and B. Remaud, J. Phys. {\bf G8}, L159 (1982).
  \bibitem{Roy84}
	  G. Royer and B. Remaud, J. Phys. {\bf G10}, 1047 (1984).
  \bibitem{Roy85}
	  G. Royer and B. Remaud, Nucl. Phys. {\bf A444}, 477 (1985).
  \bibitem{audi03}
	  G. Audia, A.H. Wapstra and C. Thibault, Nucl. Phys. {\bf A729}, 337 (2003).
  \bibitem{Mol95}
	  P. M{\"{o}}ller {\it et al.},
	  {At. Data Nucl.\ Data Tables} {\bf 59} (1995) 185.

  \bibitem{ccfullype} The code {\tt CCFULLYPE} is available from
	  \url{https://sites.google.com/site/ccfullype/}.

  \bibitem{Monta15} G. Montagnoli, A.M. Stefanini, H. Esbensen,
	  L. Corradi, S. Courtin, E. Fioretto, J. Grebosz, F. Haas,
	  H.M. Jia, C.L. Jiang, M. Mazzocco, C. Michelagnoli,
	  T. Mijatovi\'c, D. Montanari, C. Parascandolo, F. Scarlassara,
	  E. Strano, S. Szilner, D. Torresi, Phys. Lett. {\bf 746}, 300
	  (2015).
	  
  \bibitem{Stef10} A. M. Stefanini, G. Montagnoli, L. Corradi,
	  S. Courtin, E. Fioretto, A. Goasduff, F. Haas, P. Mason,
	  R. Silvestri, Pushpendra P. Singh, F. Scarlassara, and
	  S. Szilner, Phys. Rev. C {\bf 82}, 014614 (2010).
  \bibitem{Beck81} M. Beckerman, J. Ball, H. Enge, M. Salomaa,
	  A. Sperduto, S. Gazes, A. DiRienzo, and J. D. Molitoris,
	  Phys. Rev. C {\bf 23}, 1581 (1981).
  \bibitem{Beck82} M. Beckerman, M. Salomaa, A. Sperduto,
	  J.D. Molitoris, and A. DiRienzo, Phys. Rev. C {\bf 25}, 837
	  (1982).
  \bibitem{Stef09} A.M. Stefanini, G. Montagnoli, R. Silvestri,
	  L. Corradi, S. Courtin, E. Fioretto, B. Guiot, F. Haas,
	  D. Lebhertz, P. Mason, F. Scarlassara, S. Szilner,
	  Phys. Lett. {\bf B}679, 95 (2009).
 \bibitem{Esb89} H. Esbensen and F. Videbaek, Phys. Rev. C {\bf 40},
	  126 (1989).

\bibitem{Esb09} H. Esbensen and C. L. Jiang, Phys. Rev. C {\bf 79},
	  064619 (2009).
	
  \bibitem{Stef06} A. M. Stefanini, F. Scarlassara, S. Beghini,
	  G. Montagnoli, R. Silvestri, M. Trotta, B. R. Behera,
	  L. Corradi, E. Fioretto, A. Gadea, Y. W. Wu, S. Szilner,
	  H. Q. Zhang, Z. H. Liu, M. Ruan, F. Yang, and N. Rowley,
	  Phys. Rev. C {\bf 73}, 034606 (2006).
  \bibitem{das07}
	  M. Dasgupta {\it et al.}, Phys. Rev. Lett. {\bf 99}, 192701 (2007)
  \bibitem{Morton99}
	  C. R. Morton, A. C. Berriman, M. Dasgupta,
	  D. J. Hinde, and J. O. Newton, K. Hagino, and I. J. Thompson,
	  Phys. Rev. C {\bf 60} 044608 (1999).
\bibitem{Arad15} A. Shrivastava, K. Mahata, S.K. Pandit, V. Nanal,
	  T. Ichikawa, K. Hagino, A. Navin, C.S. Palshetkar,
	  V.V. Parkar, K. Ramachandran, P.C. Rout, Abhinav Kumar,
	  A. Chatterjee, and S. Kailas, in preparation.
  \bibitem{Hagino07}
	K. Hagino and Y. Watanabe, Phys. Rev. C {\bf 76}, 021601(R) (2007).

  \bibitem{Yusa10}
	S. Yusa, K. Hagino, and N. Rowley, Phys. Rev. C {\bf 82}, 024606 (2010).
  \bibitem{Yusa13}
	  S. Yusa, K. Hagino, and N. Rowley, Phys. Rev. C {\bf 88}, 044620 (2013).
	  
  \bibitem{Ack96} D. Ackermann, P. Bednarczyk, L. Corradi, D.R. Napoli,
	  C.M. Petrache, P. Spolaore, A.M. Stefanini, K.M. Varier,
	  H. Zhang, E Scarlassara, S. Beghini, G. Montagnoli,
	  L. Mtiller, G.E Segato, E Soramel, C. Signorini,
	  Nucl. Phys. {\bf A 609}, 91 (1996).

 \end{thebibliography}
\end{document}